\documentclass[10pt,conference]{IEEEtran}
\IEEEoverridecommandlockouts

\usepackage{hyperref}
\usepackage{tikz}
\usepackage{adjustbox}
\usetikzlibrary{calc,positioning}
\usepackage{booktabs}
\usepackage{tabularx}
\usepackage{array}
\usepackage{multirow}

\usepackage{booktabs}
\usepackage{multirow} 
\usepackage{pifont}
\usepackage{graphicx} % For the \resizebox command
\usepackage{makecell}
\usepackage{adjustbox}
\usepackage{forest}
\usepackage{subcaption}
\usepackage{wrapfig}
\usepackage{tablefootnote}
\usepackage{threeparttable}
\usepackage[most]{tcolorbox}

\usepackage{colortbl}
\usepackage{color}
\usepackage{xcolor}
\usepackage{makecell}
\usepackage{enumitem}
\usepackage{graphicx}
\usepackage{minted}
\usepackage{xcolor}
\usepackage{amsmath,amssymb,amsfonts}
\usepackage{algorithmic}
\usepackage{graphicx}
\usepackage{textcomp}
\usepackage{cite}
\usepackage{xcolor}
\usepackage{booktabs}
\usepackage{tabularx}
\usepackage{array}
\usepackage{microtype}
\usepackage{fontawesome}  

\makeatletter
\newcommand{\linebreakand}{%
  \end{@IEEEauthorhalign}
  \hfill\mbox{}\par
  \mbox{}\hfill\begin{@IEEEauthorhalign}
}
\makeatother

\renewcommand{\footnoterule}{%
  \kern -3pt
  \hrule width 0.35\columnwidth height 0.4pt
  \kern 2.6pt
}

\def\BibTeX{{\rm B\kern-.05em{\sc i\kern-.025em b}\kern-.08em
    T\kern-.1667em\lower.7ex\hbox{E}\kern-.125emX}}

\definecolor{lightTeal}{HTML}{add2ca}
\definecolor{softPurple}{HTML}{8c89ba}
\definecolor{paleGreen}{HTML}{e6f0db}
\definecolor{mutedOrange}{HTML}{dfa660}
\definecolor{limeGreen}{HTML}{97c04b}
\definecolor{lightBlue}{HTML}{dbeef3}
\definecolor{softRed}{HTML}{d77470}
\definecolor{dustyBlue}{HTML}{93b2d9}
\definecolor{tabletitle}{HTML}{E5DBF8} 

\definecolor{checkgreen}{HTML}{009900}
\definecolor{crossred}{HTML}{CC0000}

\AtBeginDocument{%
  \providecommand\BibTeX{{%
    Bib\TeX}}}

\definecolor{bg}{HTML}{F8F9FB}
\usepackage[framemethod=TikZ]{mdframed}
\mdfdefinestyle{MyFrame}{%
    linecolor=black,
    outerlinewidth=0.3pt,
    roundcorner=5pt,
    skipabove = 5.5pt,
    skipbelow = 5.5pt,
    innertopmargin=5.5pt, %\baselineskip,
    innerbottommargin=5.5pt, %\baselineskip,\baselineskip,
    innerrightmargin=5.5pt,
    innerleftmargin=5.5pt,
    backgroundcolor=bg, %gray!25!white
    }

\begin{document}

\title{Inside the Skill Market: From Software Engineering Activities to Reusable Agent Skills}

%\author{\IEEEauthorblockN{Anonymous Author(s)}}

\newcommand{\authorboxwidth}{4.5cm} 
\author{
\IEEEauthorblockN{Jialun Cao\IEEEauthorrefmark{1}\thanks{\IEEEauthorrefmark{1}Both authors contributed equally to this research.}}
\IEEEauthorblockA{\parbox{\authorboxwidth}{\centering
jcaoap@cse.ust.hk \\
The Hong Kong University of \\ Science and Technology\\
Hong Kong, China}}
\and
\IEEEauthorblockN{Xinru Yan\IEEEauthorrefmark{1}}
\IEEEauthorblockA{\parbox{\authorboxwidth}{\centering
yanxinru24@mails.ucas.ac.cn \\
University of Chinese Academy \\ of Sciences \\
Beijing, China}}
\and
\IEEEauthorblockN{Songqiang Chen}
\IEEEauthorblockA{\parbox{\authorboxwidth}{\centering
i9s.chen@connect.ust.hk \\
The Hong Kong University of \\ Science and Technology\\
Hong Kong, China}}
\linebreakand
\IEEEauthorblockN{Yaojie Lu}
\IEEEauthorblockA{\parbox{\authorboxwidth}{\centering
luyaojie@iscas.ac.cn \\
Institute of Software, Chinese \\
Academy of Sciences \\
Beijing, China}}
\and
\IEEEauthorblockN{Zhongxin Liu}
\IEEEauthorblockA{\parbox{\authorboxwidth}{\centering
liu\_zx@zju.edu.cn \\
Zhejiang University \\
Hangzhou, China}}
\and
\IEEEauthorblockN{Shing-Chi Cheung}
\IEEEauthorblockA{\parbox{\authorboxwidth}{\centering
scc@cse.ust.hk \\
The Hong Kong University of \\ Science and Technology\\
Hong Kong, China}}
}

\maketitle

\begin{abstract}

Software engineering (\textit{abbrev}. SE) has continuously evolved through increasingly powerful forms of reuse, from source code and libraries to components and services. Recent advances in AI agents have introduced a potentially new reusable artifact: skills. Emerging agent skill repositories and marketplaces enable developers to package, share, and reuse SE expertise as reusable skills. This trend raises a fundamental question: \textit{what SE activities are being encapsulated into reusable skills?}
Existing studies primarily focus on a broad range of skills acquisition, safety, or benchmarking, while lacking a systematic understanding of SE-specific skills and their coverage across the software development lifecycle.
To address this gap, we conduct the first large-scale empirical study of SE skills in public repositories and marketplaces. We collect and analyze a large corpus of SE skills, examining the activities they encapsulate, lifecycle coverage, evolution characteristics, and evaluation mechanisms. Our findings reveal that SE activities are increasingly becoming reusable artifacts via skills and suggest promising research opportunities for skill recommendation and engineering-oriented structuring, as well as the need for mechanisms to encapsulate high-context SE activities into reusable skills.
Overall, our study provides the first activity-centric characterization of SE skills and reveals how SE activities are increasingly being transformed into reusable skills. These findings offer new insights into skill reuse, ecosystem development, and the future of agent-centric SE.
%their composition patterns,

\end{abstract}

\section{Introduction}

Software engineering (\textit{abbrev}. SE) has long advanced through increasingly powerful forms of \textbf{reuse}~\cite{krueger1992software,frakes2005software,papazoglou2003service,wang2005component}. Over the past decades, developers have moved from reusing source code fragments to libraries, frameworks, services, and cloud-based infrastructures\cite{papazoglou2003service,wang2005component}. Each shift has changed not only how software is constructed, but also what constitutes a reusable software artifact~\cite{krueger1992software,papazoglou2003service}. Reuse has therefore remained a central mechanism for improving productivity, scalability, and knowledge transfer throughout the history of SE.

Recent advances in foundation models and AI agents are introducing a potentially new form of reuse~\cite{tang2025empowering}. 
Early work, such as Voyager~\cite{wang2023voyager}, demonstrated that \textit{the \textbf{agent skills} (\textit{abbrev}. skills), which describe specific functionality of agents, can be accumulated, retrieved, and reused} across tasks through persistent skill libraries~\cite{jiang2026sok}. Subsequently, agent frameworks such as MetaGPT~\cite{hong2024metagpt} and ChatDev~\cite{qian2024chatdev} further showed that SE processes can be \textit{structured and decomposed} through reusable workflows.

More recently, an emerging ecosystem of public skill repositories~\cite{awesome-skills,anthropic-skills} and marketplaces~\cite{liang2026skillnet,clawhub2026,skillhub2025,skillnet2026,skillsmp2025} has enabled developers to \textit{publish, discover, and share} skills for a wide range of SE tasks such as code writing~\cite{huang2023agentcoder} and testing~\cite{harman2025mutation}. Unlike traditional software artifacts, these skills encapsulate not only executable logic but also \textit{prompts, tool invocations, workflows, and domain-specific engineering knowledge}~\cite{hong2024metagpt,yang2024swe,yao2022react,schick2023toolformer,ni2026trace2skill}. 
As a result, {SE activities} that were traditionally performed manually or implemented repeatedly across projects can now be distributed as {reusable skills} and directly incorporated into agent-driven development workflows.

This emerging trend raises a fundamental question: \textbf{{what SE activities are being encapsulated into reusable skills}}? 
While recent studies have investigated how agents acquire skills,
retrieve skills, compose skills, and execute SE tasks
through agent workflows~\cite{wang2023voyager,wang2023describe,li2026skill,hong2024metagpt,yang2024swe,wang2026skillx,bouzenia2025understanding}, little is known about the activities encapsulated within these reusable skills inside the marketplace. 
There are plenty of studies on skills to either provide broad analyses across diverse domains~\cite{zhou2026comprehensive}, ranging from literature review and travel planning to domain-specific applications such as healthcare~\cite{li2026can} and finance~\cite{yu2024fincon}, or investigate specific properties such as safety~\cite{ling2026agent,xu2026agent} or benchmarking with curated evaluation suites~\cite{liu2026well,ding2026agent,zeng2025benchmarking,zhang2026coevoskills}. 

While valuable, these studies do not provide a comprehensive understanding of SE-specific skills in emerging skill markets. 
Consequently, we still lack a principled view of {\textit{which SE activities are being encapsulated into reusable skills}} and \textit{{how they are represented across the software development lifecycle}}. This gap limits our ability to reason about skills as SE artifacts: whether they systematically encode development practices such as testing, debugging, verification, or design, and whether current skill ecosystems reflect balanced coverage of the SE lifecycle or are biased toward specific types of activities, such as code generation and vulnerability detection.

To address this gap, we present the first large-scale empirical study of SE skills in popular skill marketplaces. We collect and analyze 11,497 SE skills from publicly accessible marketplaces and repositories. Through a comprehensive empirical analysis, we investigate the SE activities these skills represent, their lifecycle coverage, evolution characteristics, and evaluation mechanisms. Our \textbf{findings} reveal a notable shift in the unit of reuse: beyond reusable code and services, SE activities themselves are increasingly becoming reusable artifacts. In addition, our {findings} suggest promising research opportunities for automated skill recommendation, enhancing skill structures and version mechanisms to facilitate engineering implementation, and exploring mechanisms to build skills for SE activities in high-context lifecycle stages, such as requirement analysis and software design, shedding light on future directions for research on skill design, skill evolution management, activity reuse, and agent-centric software development.
%We further discuss the implications of this trend for future SE research, including skill quality assurance, skill evolution management, activity reuse, and agent-centric software development.
%composition patterns, 

In summary, this paper makes the following contributions:

\begin{itemize}[leftmargin=*]

\item  \textbf{A large-scale empirical study of SE-related skills as reusable artifacts of SE activities}:
We systematically collect and analyze 11,497 unique SE-related skills from public repositories and marketplaces, and provide the first activity-centric characterization of how SE tasks are encapsulated into reusable skills across the software development lifecycle. 
%The \textbf{replicable package} is available~\cite{artifact}.

\item \textbf{A reuse-oriented taxonomy and coverage analysis of SE activities in skill ecosystems}: 
We introduce a structured mapping from SE activities to reusable skills, revealing which activities are extensively supported by reusable skill abstractions and which remain underrepresented, thereby exposing the uneven reuse landscape across the SE lifecycle.

\item \textbf{An analysis of reuse effectiveness through evaluation practices and ecosystem-level skill structures}: 
We investigate how SE-related skills are evaluated and validated in practice and further examine their composition and reuse patterns within skill ecosystems, highlighting the gap between current evaluation mechanisms and the true effectiveness of reuse in real-world SE contexts.

% \item  We conduct the first large-scale empirical study of SE skills in emerging skill marketplaces.

% \item We develop a taxonomy of SE activities encapsulated by reusable skills and characterize their coverage across the software development lifecycle.

% \item We investigate the composition, evolution, and evaluation practices of SE skills, revealing key characteristics of this emerging ecosystem.

% \item We provide evidence that SE activities are increasingly being packaged as reusable skills, highlighting a new direction for software reuse in agent-driven SE.
\end{itemize}

% In summary, this paper makes the following contributions:

% \begin{itemize}[leftmargin=*]
% \item \textbf{Dataset.} We construct and release a large-scale dataset capturing the evolution history of skills, extracted from real-world repositories. This dataset provides a foundation for a wide range of empirical analyses.
% \item \textbf{In-depth Study.} We conduct a comprehensive empirical study of skill evolution, uncovering key patterns and characteristics of how skills evolve over time.
% \item \textbf{Implications.} Based on our observations of skill evolution, we derive insights into how skills should be written and maintained, and outline potential directions for future research on automated skill improvement.
% \end{itemize}

\section{Background and Related Work}

\subsection{Agent Skill Structure and Usage}

\begin{figure*}[h!]
  \centering
  \includegraphics[width=0.8\textwidth]{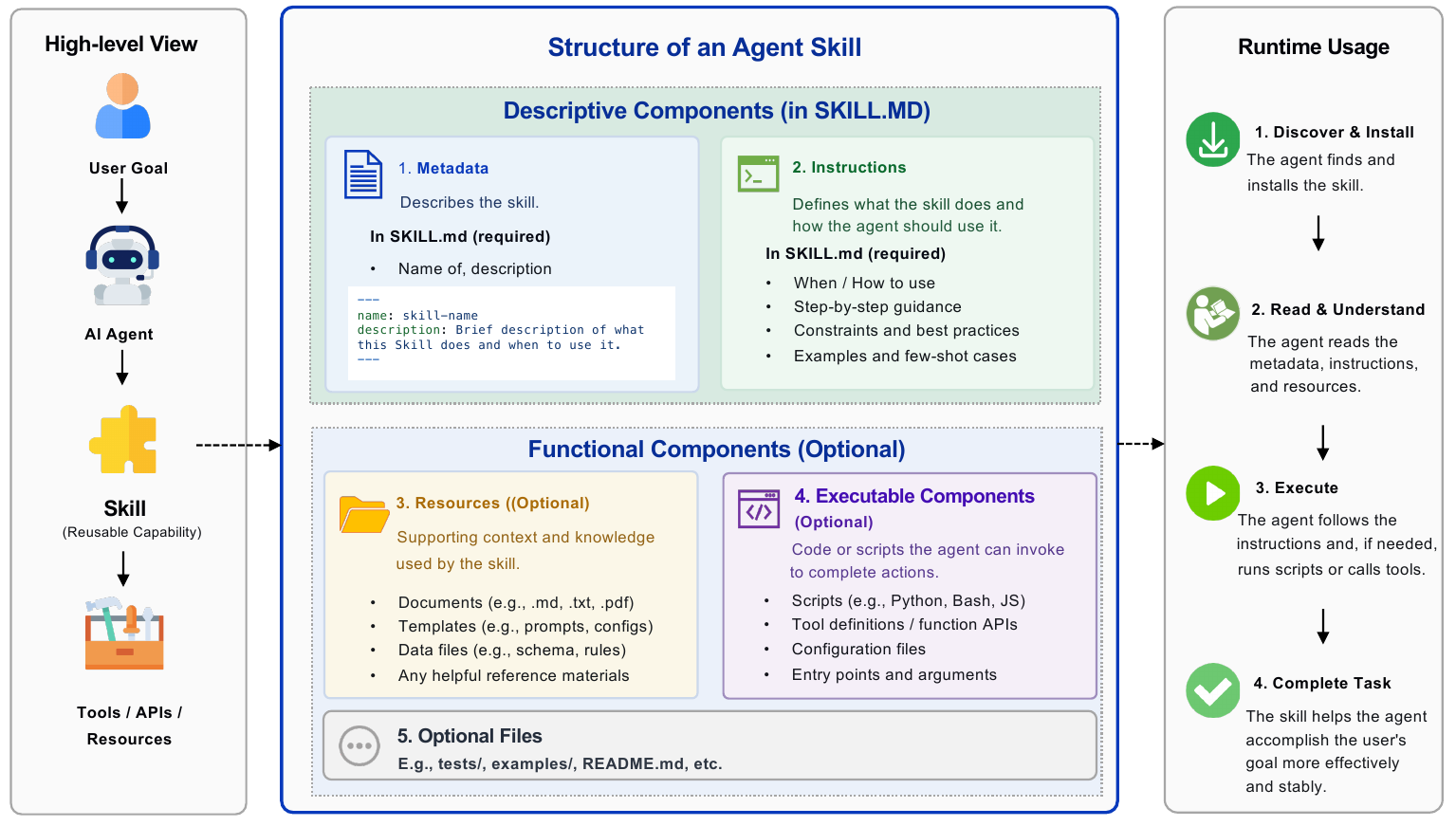}
  \caption{Structure of an Agent Skill}
  \label{fig:structure}
\end{figure*}

An agent skill, % (abbreviated as \textit{skill} in the rest of this paper), 
according to Anthropic's definition~\cite{anthropic-skills}, is a modular capability unit that extends an AI agent with reusable functionality. As shown in Figure~\ref{fig:structure}, agent skills are packaged as structured directories with standardized components and a unified execution lifecycle, enabling composability and reuse across heterogeneous agent systems. Conceptually, agent skills provide an abstraction layer between high-level user intents and low-level tools or APIs.

An agent skill consists of five components: (i) \textbf{Metadata}, defined in SKILL.md, specifying the skill name, description, and optional dependencies, versions, tags, and compatibility constraints for discovery and validation; (ii) \textbf{Instructions}, which define the semantic contract of the agent skill, including its purpose, invocation conditions, and operational steps; (iii) \textbf{Resources}, such as templates, schemas, and examples that provide contextual grounding; (iv) optional \textbf{executable components}, including scripts or tool interfaces for external system interaction; and (v) optional \textbf{supplementary files} for documentation and maintenance. The SKILL.md is mandatory, while all other components are optional.

The execution of agent skills follows a four-stage lifecycle: discovery, where relevant agent skills are selected based on metadata and task context; interpretation, where instructions and resources are parsed to determine applicability; execution, where the agent follows the skill specification and optionally invokes external components; and completion, where the agent skill contributes to final task resolution.

\subsection{Software Reuse Literature}
% Software Reuse~\cite{krueger1992software},
% Software Reuse Research: Status and Future~\cite{frakes2005software},
% Component Software~\cite{wang2005component},
% Service-Oriented Computing~\cite{papazoglou2003service}

Software reuse refers to building software systems using existing artifacts rather than developing from scratch~\cite{krueger1992software}. Early work formalizes reuse as retrieval, adaptation, and integration of reusable assets, while surveys highlight practical limitations such as integration overhead and lack of standard representations~\cite{frakes2005software}.
Recent work revisits reuse from an ecosystem and practice-oriented perspective, showing that reuse is often opportunistic and driven by ad hoc composition of existing software modules rather than planned design~\cite{makitalo2020opportunistic,capilla2019opportunities}. This shifts reuse from a static engineering activity to a dynamic system-level phenomenon. 

Beyond artifact-level reuse, recent research explores process- and execution-level reuse. Workflow systems and data pipelines enable reuse of multi-step computational procedures~\cite{cwl2021, workflow2025}, while empirical studies show that process models in practice often contain reusable behavioral structures~\cite{workflow2025}.
In parallel, software product line research advances automated and scalable reuse through variability modeling, extraction, and testing~\cite{moreira2022open,jung2024automated}. Program synthesis and API mining further extend reuse to behavioral generation, where reusable patterns are inferred from examples or codebases~\cite{gulwani2017programsynthesis,liu2019accelerating}.

More recently, the emergence of AI-native SE has further redefined reuse in the context of generative models. Recent studies~\cite{reuse-in-ai-era2025,wang2023voyager} suggest that software reuse is shifting toward LLM-based code generation and probabilistic reuse of behavioral patterns, where reuse is no longer deterministic or artifact-bound but increasingly driven by learned procedural capabilities in large-scale models.
Still, the study of reusable agent-like capabilities in AI-driven software systems is at an initial stage, which thus motivates our study.

\subsection{Surveys on Agent Skills}

Recent studies have investigated agent skills from multiple perspectives, including ecosystem-level characterization, safety analysis, benchmarking, and domain-specific applications~\cite{zhou2026comprehensive, liu2026agent, ling2026agent,xu2026agent, liu2026well,ding2026agent, zeng2025benchmarking, zhang2026coevoskills,agenticOpportunity25,li2026organizing}.
One line of work focuses on broad empirical characterization of agent skills across heterogeneous domains. These studies analyze large-scale agent skill collections spanning general-purpose and domain-specific applications, including literature assistance, travel planning, healthcare, and finance~\cite{zhou2026comprehensive, li2026can, yu2024fincon}. They primarily aim to understand distributional properties, usage patterns, and functional coverage of existing agent skill ecosystems.

Another direction studies the safety and security properties of agent skills. Existing work identifies vulnerabilities such as prompt injection and privilege escalation through systematic analysis of real-world skill repositories~\cite{ling2026agent,xu2026agent}. Complementary efforts construct benchmark suites and evaluation frameworks to assess robustness and reliability of agent skills under adversarial conditions~\cite{liu2026well, ding2026agent, zeng2025benchmarking}.
In addition, recent benchmarking-oriented studies design curated evaluation environments to measure skill quality, generalization ability, and compositional performance across tasks~\cite{liu2026well, zhang2026coevoskills}. These works emphasize controlled assessment settings rather than real-world evolution or longitudinal behavior.

Unlike prior studies, which primarily focus on broad cross-domain analyses, safety properties, or benchmark-oriented evaluations of agent skills, our study fills this gap by systematically analyzing SE-oriented agent skills in skill markets to characterize how these capabilities are represented across the software development lifecycle.
\section{Study Design}

\subsection{Research Question Design}
To understand how SE activities are transformed into reusable artifacts in emerging skill marketplaces, we structure our study around three research questions that examine skills from intrinsic properties to ecosystem-level reuse structures.

\textbf{RQ1. What are the characteristics of SE-related agent skills as reusable units of SE activities?}
As skills are increasingly positioned as reusable units of execution in agent-based systems, their fundamental design properties (\textit{e.g.}, granularity and structure) may shed light on preparation for effective new skills yet remain poorly understood. 
We thus first examine the structural and functional properties of existing SE-related skills, including their temporal growth, marketplace sources, content length, structural forms, and versioning characteristics.

\vspace{-0.15em}

\textbf{RQ2. Which SE activities are being captured and reused through skills across the software development lifecycle?}
While skills are intended to encapsulate SE capabilities, it is unclear which parts of the software development lifecycle are actually being transformed into reusable artifacts. Thus, we analyze the mapping between SE lifecycle activities and skill artifacts, identifying which types of engineering work are systematically transformed into reusable skills and how this coverage is distributed.

\textbf{RQ3. How are the SE-related agent skills evaluated and validated?}
Existing evaluation practices for skills primarily rely on task-specific success rates or benchmark performance, which may not accurately reflect reuse capability across diverse contexts. As a result, it is unclear whether current evaluation protocols measure true generalization and transferability of SE skills, or merely capture narrow, single-scenario effectiveness. This motivates us to study how skill marketplaces evaluate skills' practices, assess performance, quality, security, etc.

\begin{figure*}[t]
  \centering
  \includegraphics[width=1.0\linewidth]{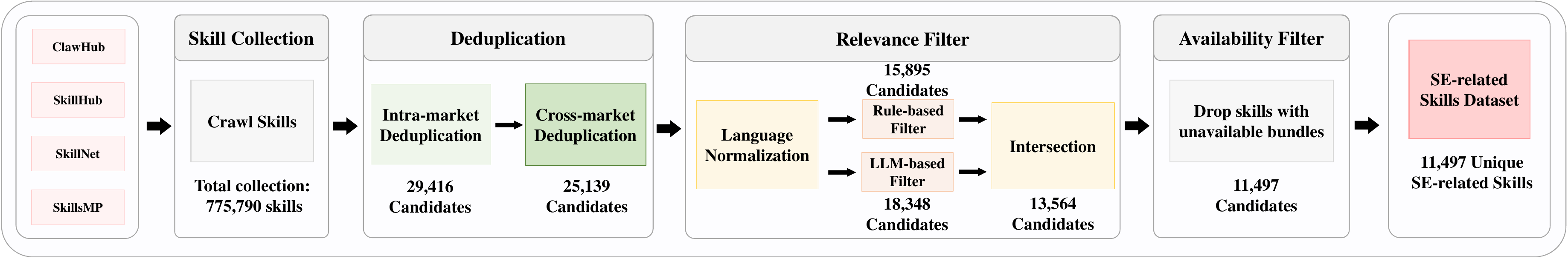}
  \caption{Collection Pipeline of SE-related Skills}
  \label{fig:2_data_collection}
\end{figure*}

\begin{figure}[t]
  \centering
  \includegraphics[width=1.0\linewidth]{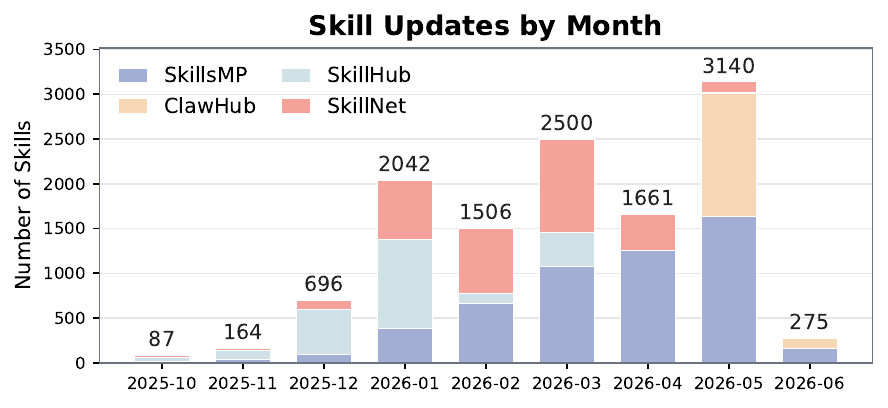}
  \caption{Statistics of Skill Updates by Month}
  \label{fig:3_timestack}
\end{figure}

\subsection{Data Collection}
We developed an automated pipeline to crawl SE-related skills from skill marketplaces (Figure~\ref{fig:2_data_collection}), and completed data collection in June 2026. Specifically, we built dedicated crawlers for four representative skill marketplaces:  \textbf{ClawHub}~\cite{clawhub2026}, \textbf{SkillHub}~\cite{skillhub2025}, \textbf{SkillNet}~\cite{skillnet2026}, and \textbf{SkillsMP}~\cite{skillsmp2025} following the selection of related work~\cite{zhou2026comprehensive}. In terms of their scope of applicable agents, ClawHub~\cite{clawhub2026} is the public registry particularly for OpenClaw~\cite{openclaw2026}, while SkillHub~\cite{skillhub2025}, SkillNet~\cite{skillnet2026} and SkillsMP~\cite{skillsmp2025} are more general-purpose, supporting Claude, Codex, Gemini, OpenCode, and other SKILL.md-compliant tools and agents.

% supports a wider range of agents, including Claude, Codex, Gemini, OpenCode, and other agents. SkillNet is open infrastructure for building and reusing AI skills, where a skill packages instructions, metadata, references, scripts, and evaluation results for later installation and reuse by agents. 
% SkillsMP is a marketplace for discovering open-source skills compatible with Claude Code, Codex, ChatGPT, and other SKILL.md-compliant tools. 
%We further curated a list of 110 software-engineering-related keywords to serve as query terms for skill crawling across these 4 platforms.

1) \textbf{Skill Collection:}
We built dedicated crawlers for four marketplaces and curated a list of 110 SE-related keywords to serve as query terms for crawling skills. 
For ClawHub, we enumerate skills by submitting query terms to the platform's search API. For each matched slug, we sequentially invoke the detail API and the download API, thereby obtaining the skill metadata together with the security evaluation results.
For SkillHub, we enumerate skills by submitting query terms to the platform's search API. For each slug returned in the search results, we then invoke the detail API to retrieve the skill metadata along with the platform's evaluation results.
For SkillNet, we query the platform's search API, performing paginated retrieval in keyword search mode to collect skill metadata and record the platform's evaluation results.
For SkillsMP, we submit query terms to the platform's search API and enumerate skills via a pagination mechanism; we then visit each skill's detail page to obtain its metadata.
In total, we collected 775,790 skills across the four marketplaces.

2) \textbf{Deduplication:} 
We first perform intra-market deduplication. For ClawHub, SkillHub, and SkillsMP, we deduplicate skills using their slug. For SkillNet, whose skill metadata does not provide a slug, we instead use the tuple of (skill name, repository URL) as the deduplication key. After this step, the total number of skills is reduced to 29,416. We perform cross-marketplace deduplication using repository URLs. After this step, we obtain 25,139 skills.

3) \textbf{Relevance Filter:}
We perform language normalization on each skill's name and description, converting all non-English text into English. We then apply a Rule-based filter. Specifically, we concatenate the skill's name and description into a single unified text and perform local rule matching against predefined lexicons. The positive lexicon spans the stages of the SE lifecycle (\textit{e.g.}, requirements, coding, testing, and deployment). The negative lexicon covers non-SE domains (\textit{e.g.}, business, academia, and healthcare). A skill is judged relevant and retained if its text matches a strong software-engineering phrase, or if the number of positive-lexicon matches reaches a predefined threshold. A skill is discarded if the negative lexicon matches dominate, or if no positive-lexicon match is found. This procedure ultimately yields 15,895 relevant skills.

We further apply an LLM-based filtering, \textit{i.e.}, for each skill, we use GPT-5.5~\cite{openai2026gpt55} to determine whether the skill was related to SE based on its name and description. 
% Specifically, for each skill, we concatenated its name and description into a unified textual input, and then used GPT-5.5~\cite{openai2026gpt55} to determine whether the skill was related to SE based on this input.
In the prompt, we anchor the judgment on the relevant and non-relevant categories of the SE lifecycle, and instruct GPT-5.5 to conservatively label a skill as non-relevant whenever the evidence is weak or the semantics are ambiguous. This step ultimately yields 18,348 relevant skills. By taking the intersection of the relevant skills identified by the two methods, {13,564 SE-relevant skills} remain.

4) \textbf{Availability Filter:}
After excluding 2,067 skills associated with deleted (404) repositories, we obtain the final dataset, comprising \textbf{11,497 unique SE-related skills}.

\begin{figure}[t]
  \centering
  \includegraphics[width=1.0\linewidth]{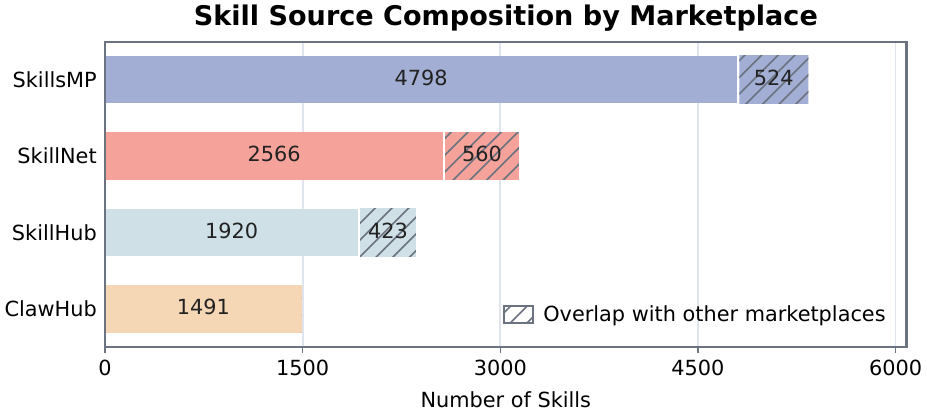}
  \caption{Statistics of Skill Source Marketplaces}
  \label{fig:4_source_stack}
\end{figure}

\section{RQ1. Characteristics of SE-related Skills}

\subsection{Skill Updates Over Time}
We counted the last update time of all SE-related  skills across four marketplaces,  shown in Figure~\ref{fig:3_timestack}. After December 2025, the number of skill updates increased significantly, reaching 2,042 in January 2026. Subsequently, from January to May 2026, it maintained an overall growth trend, indicating that the ecosystem of SE skills entered a stage of rapid growth.

% \begin{tcolorbox}[
%     colback=gray!10,
%     colframe=black,
%     boxrule=0.8pt,
%     arc=0pt,
%     left=4pt,
%     right=4pt,
%     top=3pt,
%     bottom=3pt,
%     width=\columnwidth
% ]
% \textbf{Finding 1:} The number of updates to SE-related skills increased sharply after December 2025 and maintained a growth trend.
% \end{tcolorbox}

\subsection{Skill Source}
Figure~\ref{fig:4_source_stack} shows the distribution of SE-related skills across marketplaces. SkillsMP is the largest source, contributing 5,322 skills (46.3\%), while only 722  skills (6.3\%) overlap across marketplaces, indicating both the value of multi-marketplace collection and the diversity of existing SE-related skills.

\begin{figure}[t]
  \centering
  \includegraphics[width=0.87\linewidth]{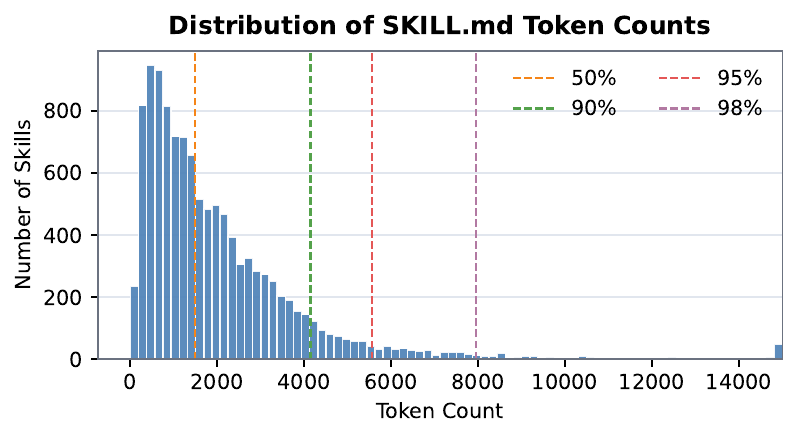}
  \caption{Distribution of SKILL.md Token Counts}
  \label{fig:5_skill_md_count}
\end{figure}

\subsection{Skill Length}
We analyze the length distribution of SE-related skill by tokenizing all SKILL.md files consistently using tiktoken~\cite{openai2026tiktoken}, OpenAI’s tokenizer, with the o200k\_base encoding~\cite{ling2026agent}. The results are shown in Figure~\ref{fig:5_skill_md_count}. Skill lengths exhibit a pronounced long tail distribution, while most skills remain relatively compact. The average skill length is 2,078 tokens. In terms of percentiles, 90\% and 95\% of skills contain no more than 4,150 and 5,565 tokens, respectively, and 98\% of skills are shorter than 7,971 tokens. The statistics indicate that most existing SE-related skills tend not to require long, verbose functionality descriptions. The top 1\% of skills exceed 11,185 tokens, with the longest skill reaching 37,499 tokens. Our manual inspection shows that such long skills often integrate multiple auxiliary materials, including reference documentation, code blocks, examples, and external API content.

% \begin{tcolorbox}[
%     colback=gray!10,
%     colframe=black,
%     boxrule=0.8pt,
%     arc=0pt,
%     left=4pt,
%     right=4pt,
%     top=3pt,
%     bottom=3pt,
%     width=\columnwidth
% ]
% \textbf{Finding 2:} Most SE-related skills are concise, whereas a small subset are substantially larger because they incorporate multiple types of auxiliary materials.
% \end{tcolorbox}

\subsection{Focus Across Skill Marketplaces}
\begin{figure}[b]
  \centering
  \includegraphics[width=0.9\linewidth]{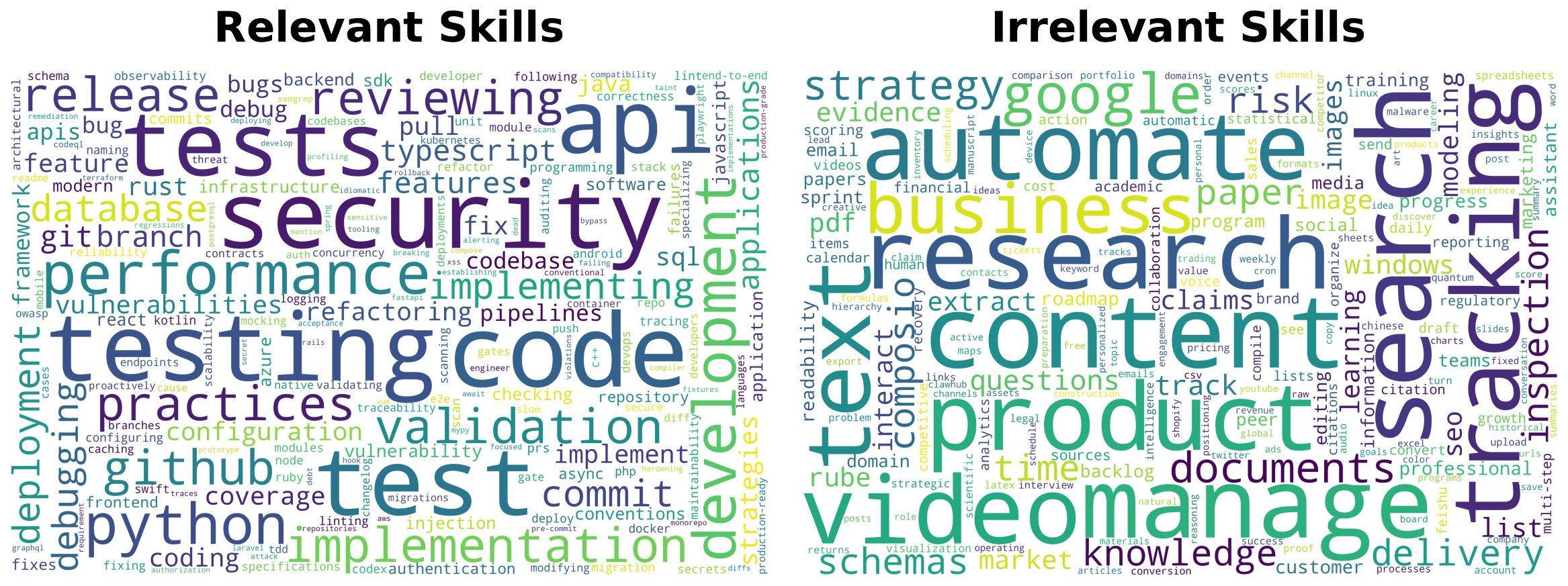}
  \caption{Word Clouds of SE-Relevant / Irrelevant Skills}
  \label{fig:6_relevant_word_cloud}
\end{figure}

We generated word clouds for relevant and irrelevant skills after the two-step filtering process, retaining only words whose frequency in one group was at least twice that in the other. Figure~\ref{fig:6_relevant_word_cloud} shows that relevant skills are dominated by SE-related terms (\textit{e.g.}, \textbf{testing, security, and code}), whereas irrelevant skills mainly focus on \textit{business, video, content management}, and \textit{research-related tasks}.
% We plotted word clouds of the descriptions of relevant and irrelevant skills after completing the two-step filtering process. For the descriptions of each skill group, a word was retained only if its frequency in the target group exceeded that in the other group by more than two times. The results are shown in Figure~\ref{fig:6_relevant_word_cloud}. Relevant skills are mainly concentrated on SE-related tasks, with prominent terms in the word cloud such as testing, security, and code, indicating that published SE-related skills currently emphasize testing, security implementation-related functionalities. In contrast, irrelevant skills are more concentrated on business, content management, and research-related tasks, as shown by the word cloud. %, with prominent terms including youtube, spreadsheet and investment.

\begin{figure}[t!]
  \centering
  \includegraphics[width=0.9\linewidth]{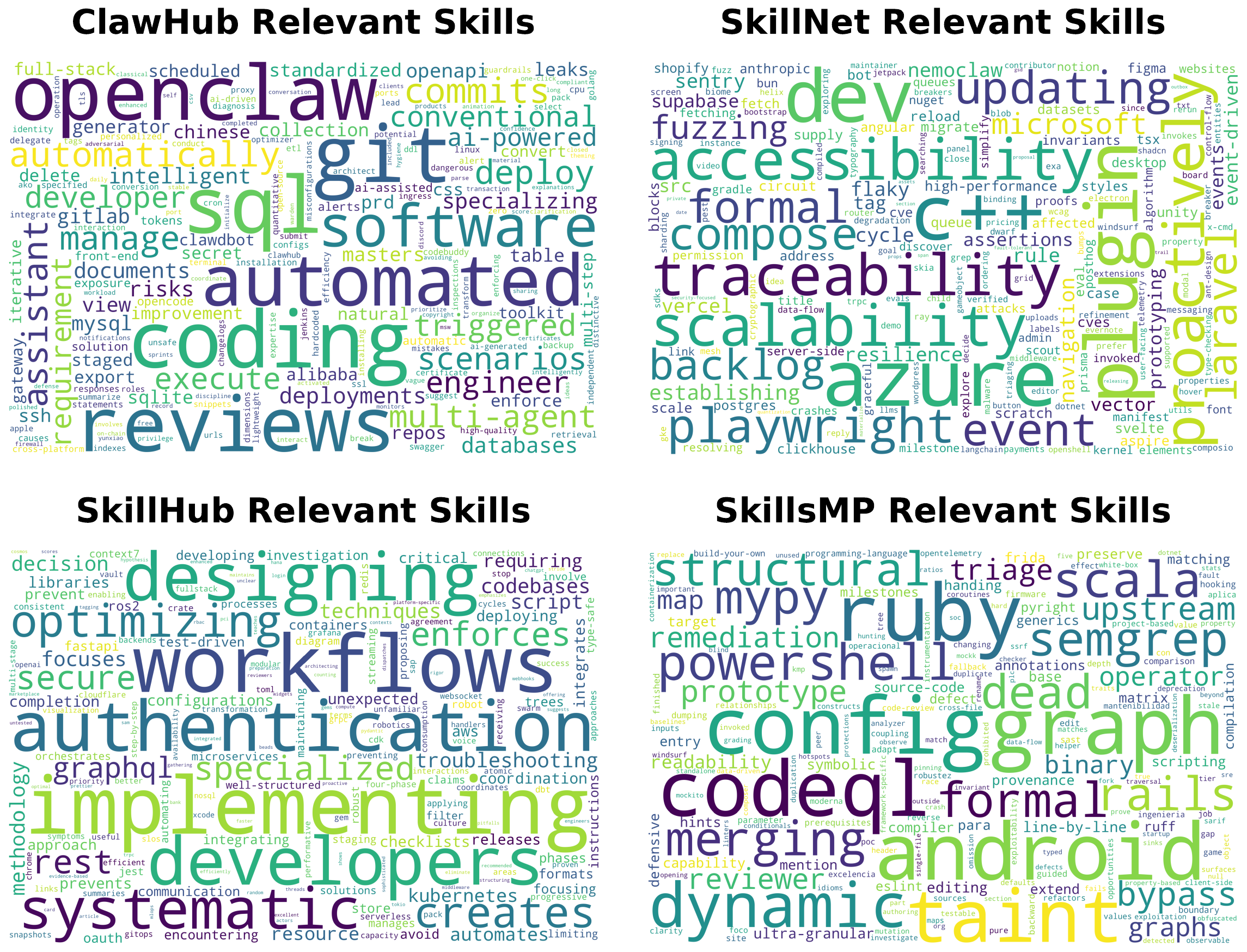}
  \caption{Word Clouds of SE-Relevant Skills in four Marketplaces}
  \label{fig:7_market_word_cloud}
\end{figure}

Figure~\ref{fig:7_market_word_cloud} plots word clouds of SE-related skills across four marketplaces: 
(1) ClawHub~\cite{clawhub2026} concentrates on coding automation and version-control workflows  (\textit{e.g.}, coding, git, sql, software, automated). 
(2) SkillNet~\cite{skillnet2026} leans toward software development and quality assurance (\textit{e.g.}, c++, azure, accessibility, traceability, plugin). 
(3) SkillHub~\cite{skillhub2025} focuses on developer workflows and software design and implementation (\textit{e.g.}, workflows, authentication, implementing, developers, designing).
(4) SkillsMP~\cite{skillsmp2025} concentrates on security analysis and configuration automation (\textit{e.g.}, ruby, config, codeql, android, powershell). 

% (1) ClawHub~\cite{clawhub2026} concentrate on coding automation and version-control workflows  (\textit{e.g.}, coding, git, sql, deploy, multi-agent). 
% (2) SkillNet~\cite{skillnet2026} lean toward software development and quality assurance practices(\textit{e.g.}, c++, azure, accessibility, traceability); 
% (3) SkillHub~\cite{skillhub2025} focus on backend service architecture and authentication mechanisms(\textit{e.g.}, implementing, workflows, authentication, secure).
% (4) SkillsMP~\cite{skillsmp2025} concentrates on security analysis and vulnerability detection (\textit{e.g.}, codeql, taint, semgrep, remediation). 

% where a word is retained only if its frequency in the target marketplace exceeds the average frequency of the other three marketplaces by more than two times. 

\begin{mdframed}[style=MyFrame]
    {\fontsize{10pt}{0}\selectfont\faStar} \textbf{Finding:} SE-related skills focus on similar core tasks (\textit{e.g.}, coding and testing) but exhibit marketplace-specific emphases on specific techniques or subjects (\textit{e.g.}, openclaw and git in ClawHub, codeql and android in SkillsMP).\\  
{\fontsize{10pt}{0}\selectfont\faLightbulbO}
\textbf{Opportunity:} 
The coexistence of shared core tasks and marketplace-specific specializations opens opportunities for \textbf{automated skill classification, routing, and retrieval}, enabling agents to select skills from the most suitable marketplace according to task characteristics.
% These findings suggest that  
% This indicates a potential ecosystem of SE-related skills focused at a high level but diverse in specific aspects like techniques, tools, and subjects, with different skills composed, linked, and organized across marketplaces or domains, motivating an ecosystem-level analysis of SE-related skills.
\end{mdframed}

\subsection{Skill Structure}
%\vspace{-0.5em}
To analyze the structural degree of skills, we conduct statistics at two levels, \textit{i.e.}, \textit{descriptive / functional} levels, according to the skill structure visualized in Figure~\ref{fig:structure}. 

\begin{figure}[t]
  \centering
  \includegraphics[width=1.0\linewidth]{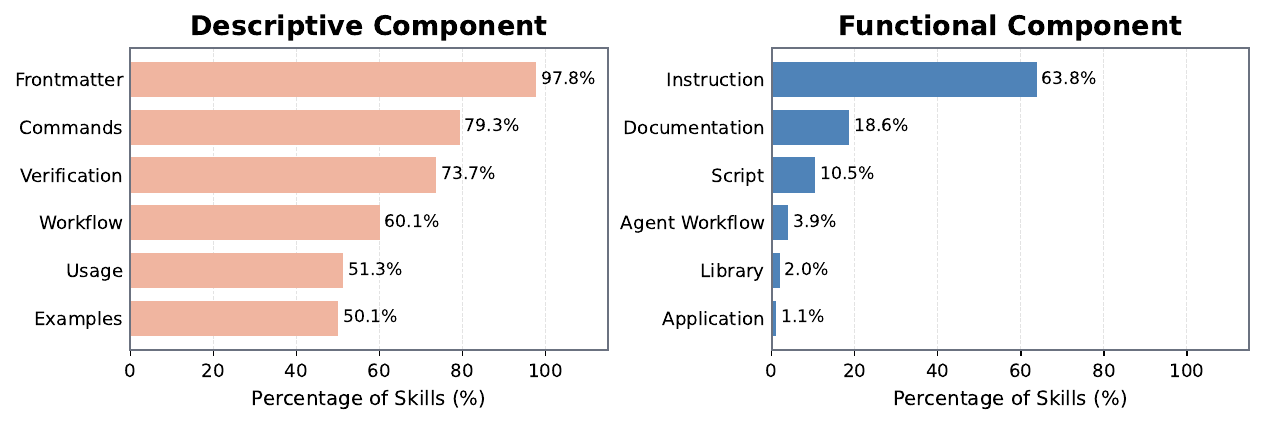}
  \caption{Distribution of Structural Characteristics}
  \label{fig:8_structure_distribution}
\end{figure}

At the \textbf{descriptive level}, we computed the coverage of key instructional elements in each skill’s SKILL.md file, covering Frontmatter, Usage, Commands, Examples, Workflow, and Verification. As shown in the left subfigure of Figure~\ref{fig:8_structure_distribution}, Frontmatter reaches 97.8\%, indicating that most SE-related skills provide metadata such as names and descriptions. The coverage rates of Commands and Verification are 79.3\% and 73.7\%, respectively, suggesting many skills describe executable operations and result verification methods. Workflow reaches 60.1\%, indicating that more than half of the skills include stepwise procedures. However, Usage and Examples account for only 51.3\% and 50.1\%, respectively, showing that nearly half of the skills still lack clear usage instructions or examples.

At the \textbf{functional level}, we identify the asset forms of each skill, including six categories: Instruction (general guidelines), Documentation (references and playbooks), Script (executable automation), Agent Workflow (agent orchestration processes), Library (reusable packages or APIs), and Application (runnable services or web apps). Specifically, for each downloaded skill bundle, we inspect its directory layout and file artifacts, such as SKILL.md, documentation files, scripts, workflow-related folders, package configurations, and service or web-application entry points. When multiple structural signals are present, we assign the skill to its primary asset form by prioritizing more engineering-oriented structures over lightweight textual ones.
The classification results are shown on the right of Figure~\ref{fig:8_structure_distribution}. Instruction accounts for the largest proportion, reaching 63.8\%, indicating that more than half of the SE-related skills are composed of instructional text without additional engineering assets. Documentation (18.6\%) ranks second, suggesting that these skills organize knowledge through documentation assets.  Only 13.6\% of skills contain code-level executable assets, including Script (10.5\%), Library (2.0\%), and Application (1.1\%). This indicates that \textbf{most SE-related skills primarily support reuse through natural-language instructions and command descriptions, rather than relying on packaged executable interfaces.}
Meanwhile,  \textbf{complex SE forms are less common, including Agent Workflow (3.9\%), Library (2.0\%), and Application (1.1\%)}. This suggests that although current skills show an emerging trend toward engineering, they still largely remain prompt or documentation-driven.

\begin{mdframed}[style=MyFrame]
    {\fontsize{10pt}{0}\selectfont\faStar} \textbf{Finding:} SE-related skills show a clear emphasis on the descriptive level, while \textbf{limited} at the functional level, especially the \textbf{engineering-oriented structures (\textit{e.g.}, scripts, library usage and workflow automation).}
% {\fontsize{10pt}{0}\selectfont\faLightbulbO}
% \textbf{Opportunity:} SE-related skills still require more support from engineering assets such as \textbf{scripts, library usage and workflow guidelines}.
\end{mdframed}

% \begin{tcolorbox}[
%     enhanced,
%     breakable,
%     colback=gray!10,
%     colframe=black,
%     boxrule=0.8pt,
%     arc=0pt,
%     left=4pt,
%     right=4pt,
%     top=3pt,
%     bottom=3pt,
%     width=\columnwidth
% ]
% \textbf{Finding 3:} SE-related skills show a clear trend toward structural organization at the descriptive level. At the functional level, skills with truly engineering-oriented structures remain limited. To evolve into mature SE components, SE-related skills still require more support from engineering assets.
% \end{tcolorbox}
%Overall, current SE-related skills show a clear trend toward structural organization at the SKILL.md level. However, at the bundle level, skills with truly engineering-oriented structures remain limited. To evolve into mature SE components, skills still require more support from engineering assets.

\begin{figure}[t]
  \centering
  \includegraphics[width=0.85\linewidth]{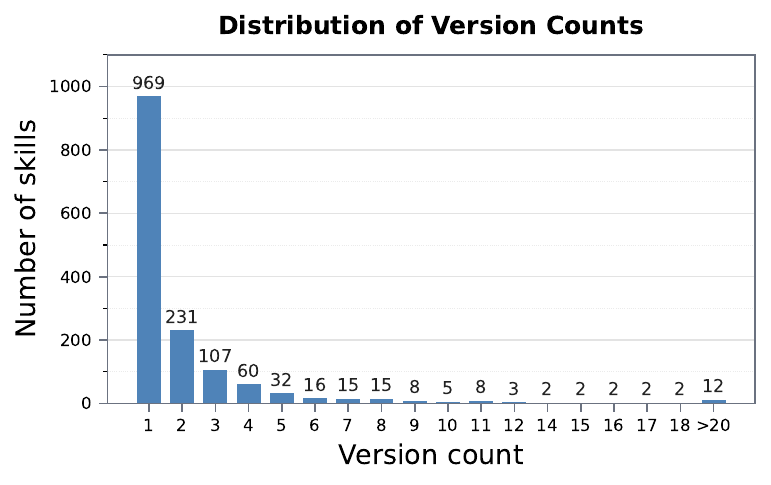}
  \caption{Distribution of Version Counts}
  \label{fig:9_version}
\end{figure}

\vspace{-1em}

\subsection{Skill Versioning}
We analyzed the number of versions for SE-related skills collected from ClawHub (Only ClawHub provides version information). As shown in Figure~\ref{fig:9_version}, most skills stopped being updated after their initial release, with 969 skills having only a single version. This indicates that version count is not a reliable indicator of skill maturity. The maximum number of versions observed is 72, and 12 skills have more than 20 versions. Manual inspection of high-version skills shows that their maintenance activities differ in nature. Some release histories correspond to genuine artifact evolution, some reflect documentation-driven hardening, while others are effectively repeated releases without substantive changes.

\begin{mdframed}[style=MyFrame]
    {\fontsize{10pt}{0}\selectfont\faStar} \textbf{Finding:} This observation suggests \textbf{a lack of indicators and standardized versioning mechanisms} for agent users to assess the maturity and evolution of SE-related skills. \\  
{\fontsize{10pt}{0}\selectfont\faLightbulbO}
\textbf{Opportunity:} Inspired by prior work~\cite{ding2026agent} which provides guidelines for skill evolution for general domains such as \textit{execution feedback} and \textit{trajectory distillation}, we highlight the {research opportunities to manage and guide SE-related skills evolution with SE methods and feedback}.
\end{mdframed}

% \begin{tcolorbox}[
%     colback=gray!10,
%     colframe=black,
%     boxrule=0.8pt,
%     arc=0pt,
%     left=4pt,
%     right=4pt,
%     top=3pt,
%     bottom=3pt,
%     width=\columnwidth
% ]
% \textbf{Finding 4:} Most SE-related skills have a small number of versions, while multi-version skills exhibit heterogeneous evolution patterns. Currently, reliable indicators of skill maturity are lacking.
% \end{tcolorbox}

\section{RQ2. Reusable SE Activity Covered by Skills}

 \begin{figure}[t]
  \centering
  \includegraphics[width=1.0\linewidth]{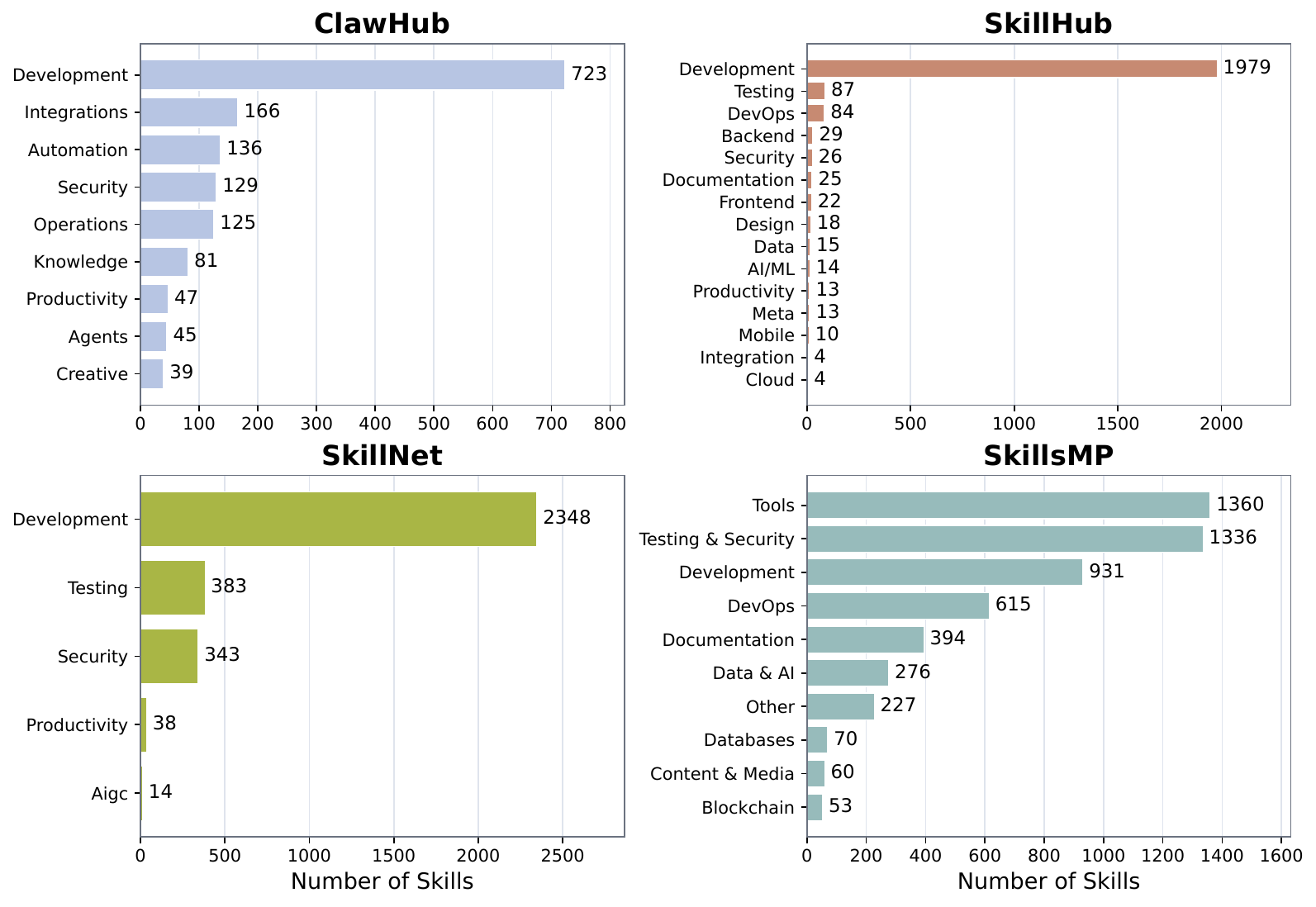}
  \caption{Distribution of Categories Across Marketplaces}
  \label{fig:9_market_category}
\end{figure}

\subsection{Marketplace Categories}
We examine the functionality distribution of SE-related skills using the category labels provided by each marketplace.
 Since different marketplaces adopt different category taxonomies, we analyze each marketplace separately. For SkillHub, we normalize categories according to its official website categories. For SkillsMP, we map fine-grained categories into major categories. Since ClawHub updated its category taxonomy after our initial data collection, we re-crawled the categories of the collected ClawHub skills on June 20, 2026.

Figure~\ref{fig:9_market_category} shows the category distribution across the four marketplaces.  In ClawHub, Development is the largest category with 723 skills, followed by Integrations, Automation, Security, and Operations. SkillHub shows a stronger concentration: Development accounts for 1,979 skills, substantially exceeding all other categories such as Testing, DevOps, and Backend. Similarly, SkillNet is dominated by Development with 2,348 skills, followed by Testing and Security. In contrast, SkillsMP exhibits a more balanced distribution, with Tools, Testing \& Security, Development, and DevOps forming the major functional groups.
These results indicate that skill marketplaces now provide a wide range of reusable SE capabilities, with development as the dominant category. Testing, Security, Development, and DevOps also attract more attention than other areas. The distribution suggests that the current marketplace supply is skewed toward certain SE activities rather than all software lifecycle activities.
At the same time, we observed differences among marketplace taxonomies. Although the varying native marketplace categories are useful for understanding platform-specific positioning, they may hinder cross-marketplace comparison. This also motivates the construction of a unified task taxonomy in the following analysis in this RQ.

% These results indicate that current SE-related skills are primarily designed to support development activities, while testing, security, automation, and DevOps also represent important functional areas. At the same time, the differences among marketplace taxonomies suggest that native marketplace categories are useful for understanding platform-specific positioning, but they are insufficient for cross-marketplace comparison. This motivates the construction of a unified task taxonomy in the following analysis.

\subsection{SE Lifecycle Coverage}\label{sec:lifecycle}
To analyze the coverage of relevant skills across the SE lifecycle, we map each skill to a unified lifecycle taxonomy.
We define eight lifecycle stages: Requirement, Plan \& Design, Implementation, Code Review, Testing, Release, Deployment, and Maintenance \& Operations. We use Qwen3.6-35B-A3B~\cite{qwen36_35b_a3b} to annotate each skill. For every skill, we construct an input instance consisting of its name, description, and SKILL.md content. The prompt provides definitions of the eight lifecycle stages and requires the model to return exactly one label from the predefined taxonomy. For skills that span multiple lifecycle stages, the model is instructed to identify the stage that reflects the skill’s dominant reusable capability.

\begin{figure}[t]
  \centering
  \includegraphics[width=1\linewidth]{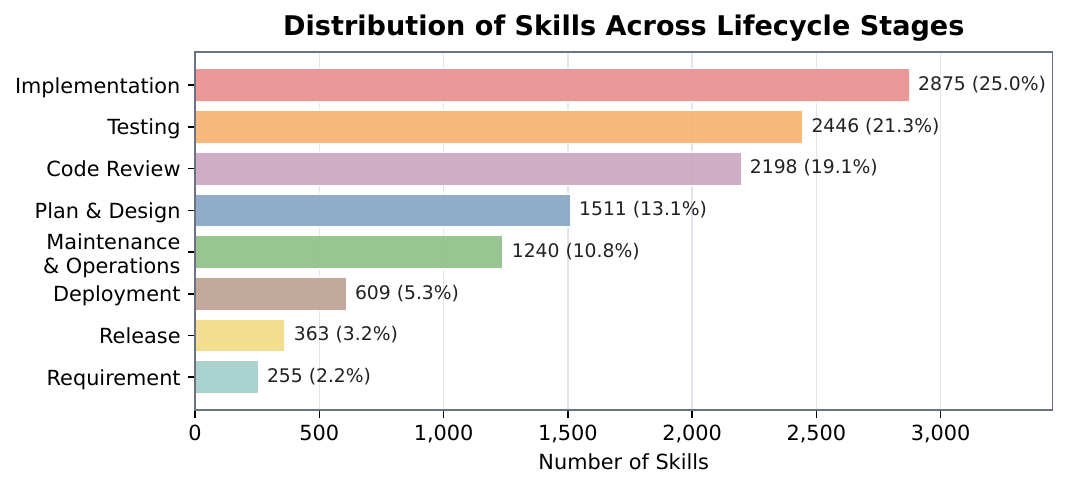}
  \caption{Distribution of Skills Across Lifecycle Stages}
  \label{fig:10_lifecycle}
\end{figure}

Figure~\ref{fig:10_lifecycle} reports the lifecycle distribution of the SE-related skills. The results show a clear imbalance across the software lifecycle. Implementation is the largest category, containing 2,875 skills (25.0\%), followed by Testing with 2,446 skills (21.3\%) and Code Review with 2,198 skills (19.1\%). Together, these three stages account for 65.4\% of all SE-related skills. This indicates that \textbf{code-centric development and verification activities}, with implementation, testing, and code review form the current dominant lifecycle stages.
By contrast, earlier and later lifecycle stages receive substantially less support. 
Plan \& Design contains 1,511 skills (13.1\%), while Maintenance \& Operations, Deployment account for 1,240 (10.8\%), and 609 (5.3\%) skills, respectively. Release and Requirement are the least represented stages, with only 363 (3.2\%) and 255 (2.2\%) skills.

Overall, the imbalance across lifecycle stages suggests that SE-related skills are easier to form around stages with \textbf{standardized procedures and relatively objective feedback}. For example, Implementation, Testing, and Code Review are typically grounded in concrete artifacts and can be evaluated through executable results, static findings, or localized code-quality judgments. This suggests practitioners leverage abundant existing skills or distill new reusable skills when working on such stages.
By contrast, Requirement, Release, and Maintenance \& Operations depend more on project-specific context, runtime environments, and subjective success criteria, which increases the difficulty of reusable abstraction.

% \begin{tcolorbox}[
%     colback=gray!10,
%     colframe=black,
%     boxrule=0.8pt,
%     arc=0pt,
%     left=4pt,
%     right=4pt,
%     top=3pt,
%     bottom=3pt,
%     width=\columnwidth
% ]
% \textbf{Finding 5:} 
% \end{tcolorbox}

\begin{mdframed}[style=MyFrame]
    {\fontsize{10pt}{0}\selectfont\faStar} \textbf{Finding:} SE-related skills are unevenly distributed, with rich support for implementation, testing, and code review stages, but \textbf{limited support for requirements, planning, deployment and maintenance}. \\
{\fontsize{10pt}{0}\selectfont\faLightbulbO}
\textbf{Opportunity:} The uneven lifecycle coverage reveals substantial opportunities to encapsulate yet-to-be-encapsulated reusable SE activities into reusable skills. Their lower coverage reflects not only a gap in skill quantity in the wild, but also a deeper challenge: \textbf{making high-context lifecycle work reusable without stripping away the context that makes it meaningful.}
\end{mdframed}

\begin{figure}[t]
  \centering
  \includegraphics[width=1\linewidth]{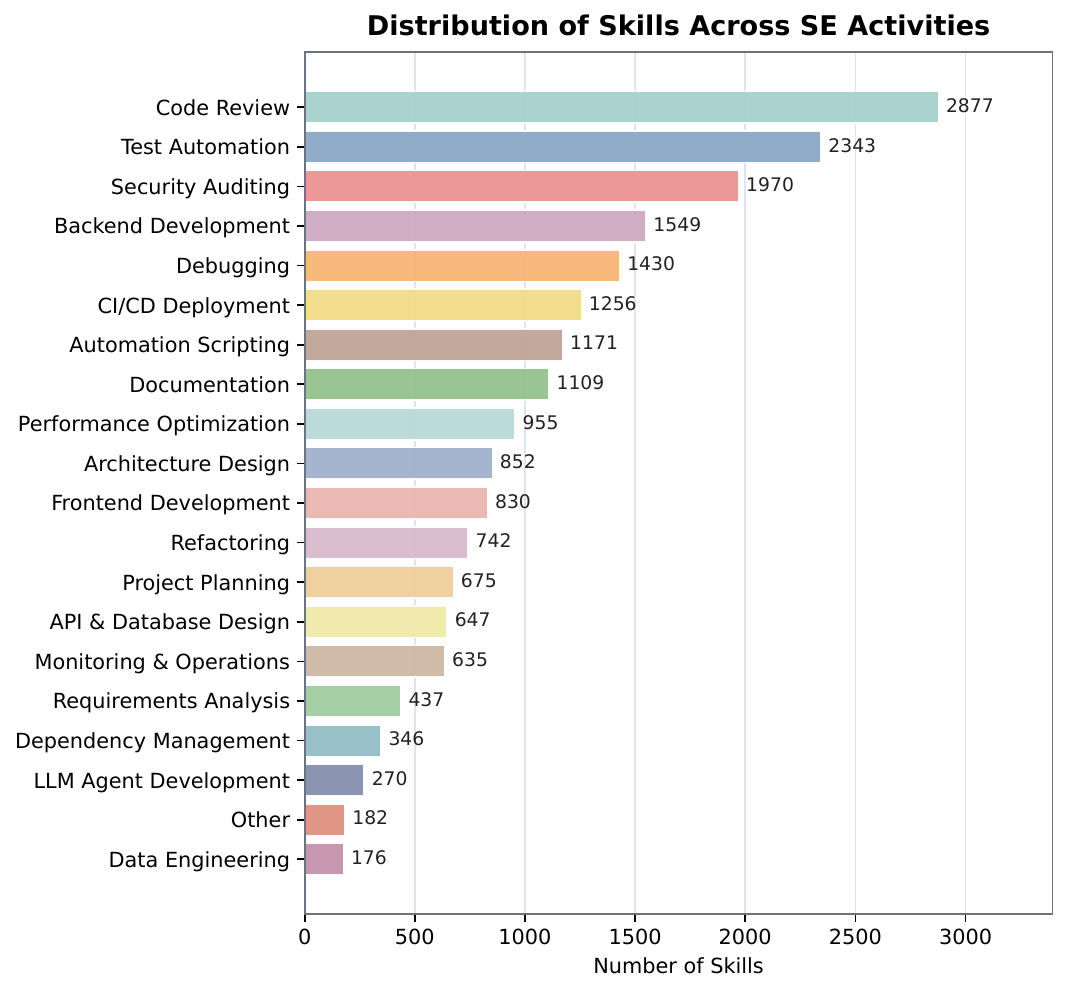}
  \caption{Distribution of Skills Across SE Activities}
  \label{fig:11_task}
\end{figure}

\vspace{-0.9em}

\subsection{SE Activity Coverage}
\label{sec:SE_related_Activity}

To further understand which SE activities are covered by existing SE-related skills, we analyze these skills at the level of fine-grained SE activities.
We construct a unified taxonomy containing twenty SE activities, including Code Review, Test Automation, Security Auditing, Backend Development, Debugging, CI/CD Deployment, Automation Scripting, Documentation, Performance Optimization, Architecture Design, Frontend Development, Refactoring, Project Planning, API \& Database Design, Monitoring \& Operations, Requirement Analysis, Dependency Management, LLM Agent Development, Data Engineering and Other (\textit{e.g.}, documentation refinement). We use Qwen3.6-35B-A3B as the annotator. For each skill, we provide the model with the skill name, description, and the content of its SKILL.md. For each of the twenty activities, the model independently judges whether the skill encapsulates that activity.
%~\cite{hou2024large}
The results are shown in Figure~\ref{fig:11_task}.
Overall, Code Review is the most common activity, with 2,877 skills, followed by Test Automation with 2,343 skills and Security Auditing with 1,970 skills. Together, these three activities account for 35.5\% of all task assignments. In contrast, Dependency Management appears in 346 skills, LLM Agent Development in 270 skills, and Data Engineering in 176 skills.

These activity-level results echo the previously identified imbalanced skill coverage across lifecycle stages (Section~\ref{sec:lifecycle}). The highly covered activities (\textit{e.g.}, \textbf{Code Review, Test Automation, Security Auditing, and Debugging}) can \textbf{often be decomposed into repeatable procedures} with relatively clear inputs, outputs, and evaluation signals, which facilitates their encapsulation as reusable skills. Meanwhile, the less covered activities like data engineering, requirements analysis, and project planning are \textbf{more shaped by domain assumptions, system history, and human preferences}. They are harder to encapsulate as reusable skills \textbf{since they require not only procedural automation, but also the ability to recover and adapt to project-specific context}.

\begin{figure}[t]
  \centering
  \includegraphics[width=1.0\linewidth]{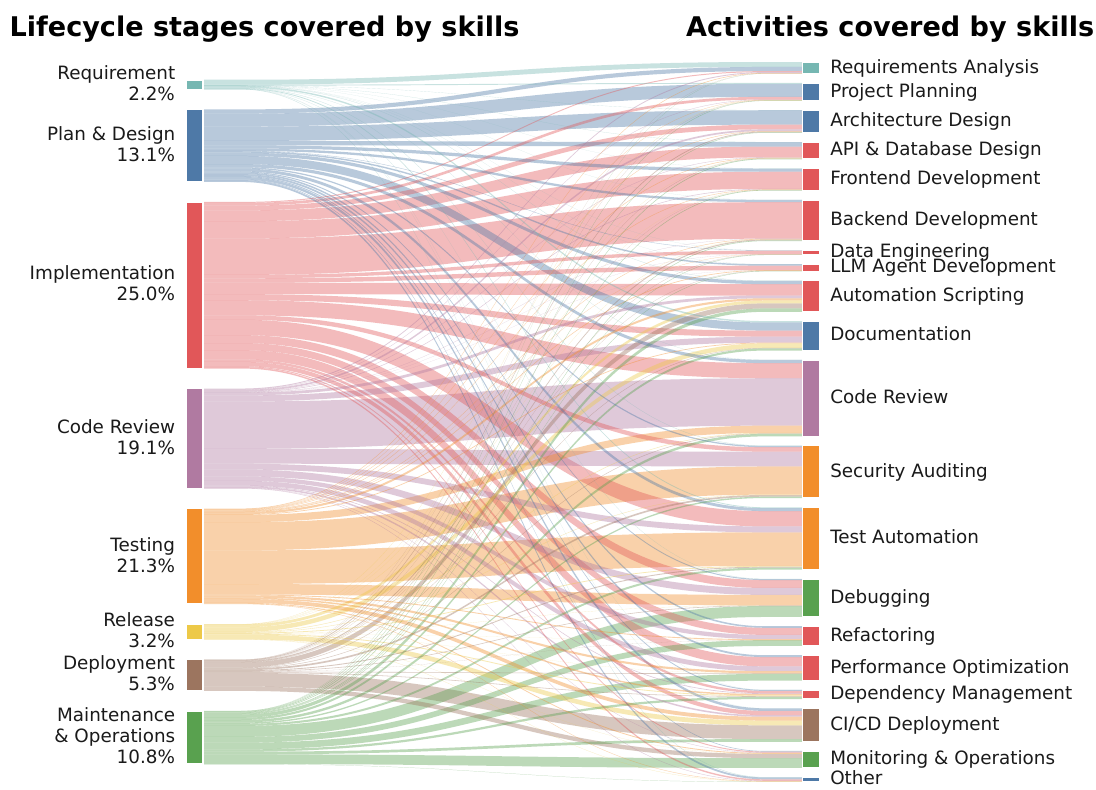}
  \caption{Lifecycle Stage to SE Activity Mapping of Skills}
  \label{fig:lifecycle-and-activity-mapping}
\end{figure}

% \begin{tcolorbox}[
%     colback=gray!10,
%     colframe=black,
%     boxrule=0.8pt,
%     arc=0pt,
%     left=4pt,
%     right=4pt,
%     top=3pt,
%     bottom=3pt,
%     width=\columnwidth
% ]
% \textbf{Finding 6:} SE-related skills show stronger coverage of code-facing and verification-oriented activities, while context-dependent and infrastructure-specific tasks remain comparatively sparse.
% \end{tcolorbox}

\begin{mdframed}[style=MyFrame]
    {\fontsize{10pt}{0}\selectfont\faStar} \textbf{Finding:} Current skills show stronger coverage of code-facing and verification-oriented activities, while context-dependent and infrastructure-specific tasks remain sparse.\\
{\fontsize{10pt}{0}\selectfont\faLightbulbO}
\textbf{Opportunity:} How to encapsulate context-dependent SE activities into reusable skills without sacrificing essential context through \textbf{abstraction} deserves investigation.
\end{mdframed}

\subsection{Mapping Between Lifecycle Stages and SE Activities}

Beyond individual distributions, we further examine the mapping between lifecycle stages and SE activities encapsulated in skills. Figure~\ref{fig:lifecycle-and-activity-mapping} shows the mapping distribution. We can observe that most SE activities are \textbf{strongly associated with a single lifecycle stage} rather than being evenly distributed across multiple stages. Implementation-related activities, including backend development, frontend development, API and database design, automation scripting, and LLM agent development, overwhelmingly originate from the \textbf{Implementation} stage. Likewise, activities such as \textbf{Code Review}, \textbf{Test Automation}, and \textbf{Security Auditing} are almost exclusively associated with their corresponding lifecycle stages. This indicates that reusable skills are generally organized around stage-specific tasks instead of cross-lifecycle capabilities.

\begin{mdframed}[style=MyFrame]
    {\fontsize{10pt}{0}\selectfont\faStar} \textbf{Finding:} Most SE activities are strongly associated with a single lifecycle stage rather than crossing multiple stages. \\
{\fontsize{10pt}{0}\selectfont\faLightbulbO}
\textbf{Opportunity:} Developing reusable skills that span multiple lifecycle stages could enable more {seamless end-to-end SE workflows}.
\end{mdframed}

\vspace{-0.5em}

In the lens through lifecycle stages, stages do not exhibit a one-to-one correspondence with SE activities. Particularly, \textbf{Implementation} (25.0\%) covers the broadest range of SE activities. Besides traditional coding tasks, it also includes Debugging, Documentation, Refactoring, Dependency Management, CI/CD Deployment, Performance Optimization and Project Planning. This suggests that Implementation has become the primary stage where reusable engineering procedures are encapsulated into skills.

Similarly, \textbf{Plan \& Design} (13.1\%) also distributes its skills across a diverse set of activities in practice, including Project Planning, Architecture Design, Requirements Analysis,  Documentation, API \& Database Design, and even implementation-related tasks. Rather than corresponding to one dominant activity, it functions as a transitional stage that decomposes high-level design decisions into multiple more concrete downstream engineering activities.

\begin{mdframed}[style=MyFrame]
    {\fontsize{10pt}{0}\selectfont\faStar} \textbf{Finding:} Implementation is the most diversified stage, followed by Plan \& Design stage, which serves as a bridge rather than a single activity category.\\
{\fontsize{10pt}{0}\selectfont\faLightbulbO}
\textbf{Opportunity:} For skills for the stage of \textbf{Implementation}, one may consider \textbf{decomposing complex implementation workflows into modular and composable reusable skills} to facilitate skill reuse and orchestration. 
For \textbf{Planing \& Design}, one may consider \textbf{explicitly modeling the design rationale and decision propagation}, enabling planning knowledge to be reused throughout subsequent stages.
\end{mdframed}

%\vspace{-0.5em}

On the contrary, \textbf{Deployment} (5.3\%) and \textbf{Maintenance \& Operations} (10.8\%) remain specialized, mainly encapsulating operational activities such as CI/CD Deployment, Monitoring \& Operations and Debugging, with relatively limited connections to upstream SE activities. Compared with implementation and testing, these lifecycle stages exhibit a much narrower activity spectrum.

\begin{mdframed}[style=MyFrame]
{\fontsize{10pt}{0}\selectfont\faStar} \textbf{Finding:} Taken together, the mapping between lifecycle stages and SE activities is inherently many-to-many rather than one-to-one. \\
{\fontsize{10pt}{0}\selectfont\faLightbulbO}
\textbf{Opportunity:} 
\textbf{Skills naturally transcend traditional lifecycle boundaries}. As agents increasingly orchestrate reusable engineering capabilities, the \textbf{conventional SE lifecycle may need to be revisited and reorganized around reusable activities}.
\end{mdframed}

\section{SE-related Skill Evaluation}

\begin{table}[t]
\centering
\scriptsize
\renewcommand{\arraystretch}{1.15}
\caption{Evaluation Metrics for Skills}
\label{tab:evaluation_metrics}
\begin{tabularx}{\linewidth}{@{}p{1.45cm}@{}p{2.15cm}@{ }X@{}}
\toprule
\textbf{Marketplace} & \textbf{Metric} & \textbf{Description} \\
\midrule

\multirow[c]{12}{1.45cm}{ClawHub}
& VirusTotal & Assesses whether the skill contains known malicious artifacts or security risks~\cite{openclaw_security_audits}. \\
& SkillSpector & Identifies potential vulnerabilities, malicious patterns, and security risks~\cite{nvidia_skillspector}. \\
& Purpose \& Capability & Assesses alignment between the skill's claimed purpose and actual capability~\cite{openclaw_security_prompt}. \\
& Instruction Scope & Evaluates whether instructions remain within the boundaries of the stated purpose~\cite{openclaw_security_prompt}. \\
& Install Mechanism & Assesses security risks associated with installation content and mechanisms~\cite{openclaw_security_prompt}. \\
& Credentials & Evaluates whether the secrets and environment access requested are proportionate~\cite{openclaw_security_prompt}. \\
& Persistence \& Privilege & Assesses whether requested persistence and privileges exceed reasonable boundaries~\cite{openclaw_security_prompt}. \\

\midrule

\multirow[c]{7}{1.45cm}{SkillHub}
& Security & Assesses security properties of the skill~\cite{skillhub2025}. \\
& Practicality & Evaluates practical utility of the skill~\cite{skillhub2025}. \\
& Output Quality & Assesses output quality produced by the skill~\cite{skillhub2025}. \\
& Instruction Clarity & Evaluates clarity of the skill instructions~\cite{skillhub2025}. \\
& Maintainability & Assesses maintainability of the skill~\cite{skillhub2025}. \\
& Innovation & Evaluates degree of innovation demonstrated by the skill~\cite{skillhub2025}. \\

\midrule

\multirow[c]{9}{1.45cm}{SkillNet}
& Safety & Assesses operational risks and robustness against prompt injection or adversarial manipulation~\cite{liang2026skillnet}. \\
& Completeness & Evaluates whether critical steps, prerequisites, dependencies, and constraints are sufficiently specified~\cite{liang2026skillnet}. \\
& Executability & Assesses whether the skill can be reliably executed in sandboxed agent environments~\cite{liang2026skillnet}. \\
& Cost Awareness & Quantifies execution overhead in terms of latency, computation, and API usage costs~\cite{liang2026skillnet}. \\
& Maintainability & Evaluates modularity and updateability without disrupting dependencies or compatibility~\cite{liang2026skillnet}. \\
\bottomrule
\multicolumn{3}{l}{* The explanation is not explicitly provided by the marketplace, so we explained} \\
\multicolumn{3}{l}{according to the literal meaning.}
\end{tabularx}
\end{table}

\begin{figure}[t]
  \centering
  \includegraphics[width=0.8\linewidth]{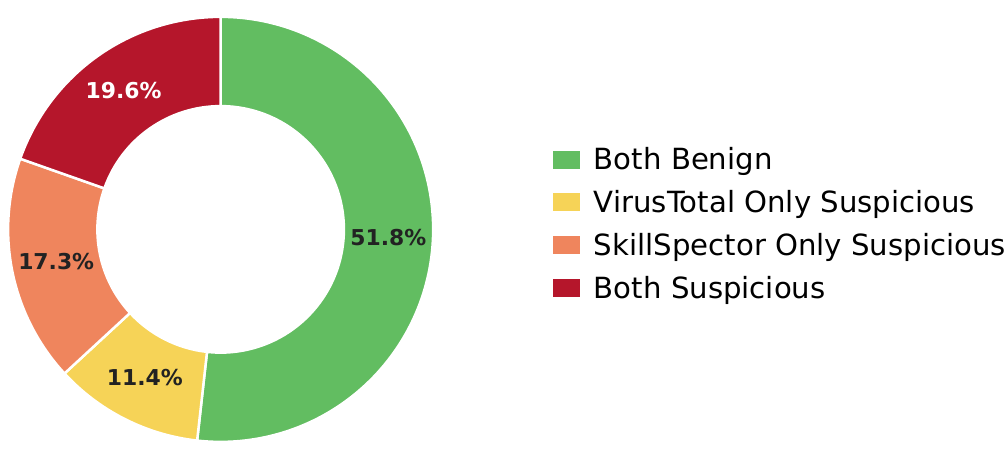}
  \caption{ClawHub SE-related Skills Security Analysis Results}
  \label{fig:14_clawhub_results}
\end{figure}

\subsection{Evaluation Metrics Across Marketplaces}
To understand how SE-related skills are currently validated by marketplaces before reuse, we first compile the evaluation metrics of each marketplace based on its public documentation and the metadata of its skills (Table~\ref{tab:evaluation_metrics}). 
Across marketplaces, evaluation metrics emphasize different aspects of skills. 
For ClawHub, the evaluation is primarily oriented toward security and boundary control: its metrics combine signature-based scanning, including VirusTotal, an online malware and URL scanning platform commonly used for malware analysis, threat intelligence, and security dataset labeling~\cite{virustotal}, and SkillSpector, NVIDIA's open-source AI skill security scanner for detecting malicious patterns, hidden instructions, excessive permissions, prompt injection, and data leakage risks before skill installation~\cite{nvidia_skillspector}. 
In addition, ClawHub uses five LLM-judged boundary checks: Purpose \& Capability, Instruction Scope, Install Mechanism, Credentials, and Persistence \& Privilege~\cite{clawhub2026}.
SkillHub adopts a broader general-quality and utility-oriented evaluation scheme, reporting six dimensions: Security, Practicality, Output Quality, Instruction Clarity, Maintainability, and Innovation~\cite{skillhub2025}. 
SkillNet focuses more on execution-oriented readiness, proposing a multi-dimensional evaluation framework that covers Safety, Completeness, Executability, Cost Awareness, and Maintainability~\cite{liang2026skillnet}. 
SkillsMP does not provide evaluation metadata.

\begin{figure}[t]
  \centering
  \includegraphics[width=1\linewidth]{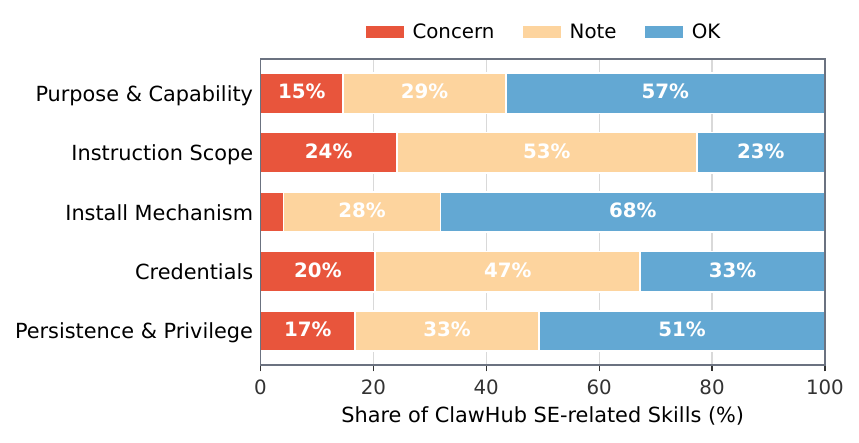}
  \caption{ClawHub SE-related Skills LLM Evaluation Results}
  \label{fig:15_clawhub_results}
\end{figure}

\subsection{Marketplace-Native Evaluation}
We examine the marketplace-reported evaluation results associated with SE-related skills. The following provides a detailed analysis of the evaluation results for each marketplace:

\textbf{1) ClawHub\footnote{Following an update to ClawHub's evaluation metrics, we re-crawled the evaluation fields for ClawHub skills on June 26, 2026.
% Because ClawHub updated its evaluation metrics after our initial data collection, we re-crawled the evaluation-related fields for ClawHub-sourced skills on June 26, 2026.
}:} 
Figure~\ref{fig:14_clawhub_results} compares the
results of VirusTotal and SkillSpector on ClawHub SE-related skills. 
Although the largest group of skills is marked benign by both scanners, a substantial proportion still triggers at least one suspicious signal: 19.6\% are marked suspicious by both scanners, 11.4\% are flagged only by VirusTotal, and 17.3\% only by SkillSpector. 
Our manual inspection of the one-sided suspicious cases further shows that the two scanners capture different types of risk. 
VirusTotal-only cases are mainly associated with implementation-level security concerns, such as shell or system-command execution, unsanitized inputs, remote script execution, supply-chain-style installation patterns, or access to API tokens and cloud credentials. 
In contrast, SkillSpector-only cases more often reflect agent-skill-specific boundary risks, including over-broad instruction scope, unclear least-privilege declarations, credential exposure, and persistence or privilege concerns. 

Figure~\ref{fig:15_clawhub_results} further reports ClawHub's five LLM-judged boundary checks, each labeled OK, Note, or Concern. Among the five checks, Install Mechanism shows the most favorable result, with 68\% of skills rated as OK, indicating that most ClawHub SE-related skills do not expose obvious installation-level risks. 
%Purpose \& Capability and Persistence \& Privilege have 57\% and 51\% of skills rated as OK, respectively, while 15\% and 17\% are flagged as Concern. 
However, for Instruction Scope, only 23\% of skills are rated as OK, while 53\% are marked as Note and 24\% as Concern. 
For Credentials, only 33\% of skills are rated as OK, and 20\% are flagged as Concern. 
These results indicate that many ClawHub SE-related skills still lack clearly constrained instructions or reasonable access-permission requirements, which may limit the safety and transferability of skill reuse. 
Overall, ClawHub's evaluation reveals security and boundary-control issues, but \textit{mainly captures prerequisites for safe reuse rather than directly validating cross-context reuse capability}.

\begin{figure}[t]
  \centering
  \includegraphics[width=1\linewidth]{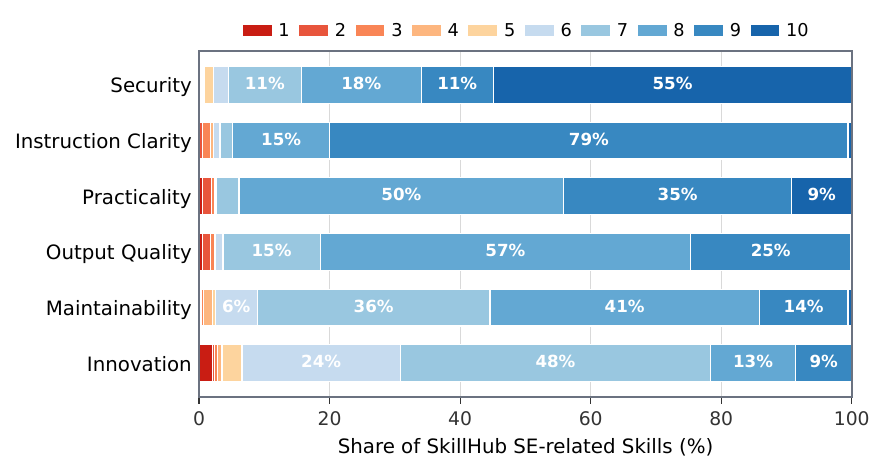}
  \caption{SkillHub SE-related Skills Evaluation Results}
  \label{fig:16_skillhub_results}
\end{figure}

\textbf{2) SkillHub:}
Figure~\ref{fig:16_skillhub_results} reports the score distribution of SkillHub SE-related skills across six marketplace-native evaluation
dimensions. 
Overall, the scores are concentrated in the upper range, suggesting that SkillHub SE-related skills are generally rated as usable under the platform's own evaluation criteria.
Innovation is the weakest dimension, with 6\% of skills receiving scores of 5 or below.
We manually inspected the \textit{cons (limitations)} associated with skills receiving low innovation scores and found that their lower innovation is primarily attributable to placeholder-style content, limited implementation depth and automation, dependence on external documentation, generic best-practice-oriented guidance, and restricted applicability to specific project contexts.
However, this concentration may also reflect limited discriminative power in distinguishing skills with different levels of quality or reuse readiness. 
Meanwhile, these evaluation metrics are mainly \textit{limited to the static quality} of the skill artifact itself, and cannot assess its transferability and generalizability dynamically.

\begin{figure}[t]
  \centering
  \includegraphics[width=0.85\linewidth]{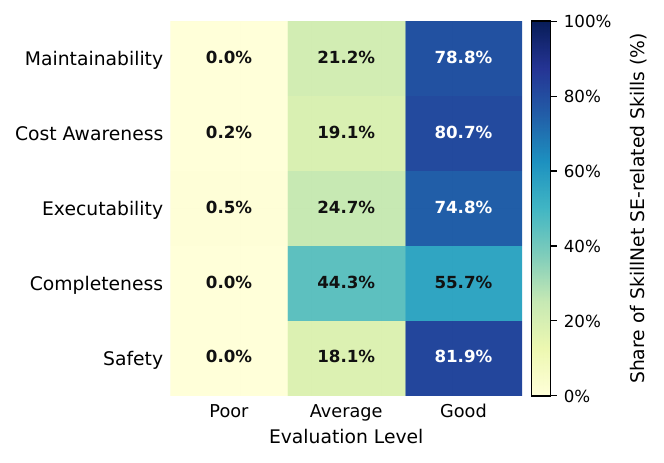}
  \caption{SkillNet SE-related Skills Evaluation Results}
  \label{fig:17_skillnet_results}
\end{figure}

\textbf{3) SkillNet:}
SkillNet provides a three-level evaluation (\textit{i.e.}, Poor, Average, and Good) over five dimensions.
Figure~\ref{fig:17_skillnet_results} shows that Good and Average ratings dominate across all dimensions, while Poor ratings are rare.
In the Safety dimension, 81.9\% of skills received a Good rating and 18.1\% received an Average rating, with no skills rated as Poor.
In Cost Awareness, 0.2\% of skills were rated as Poor. Manual inspection of the per-metric \textit{reason} fields in metadata indicates that the cases rated as Poor are mainly penalized because they prescribe open-ended, resource-intensive workflows (\textit{e.g.}, repeated test execution, exhaustive codebase tracing, agent orchestration), without explicit mechanisms to bound runtime, API calls, model usage, or infrastructure cost.
In the Executability dimension, 0.5\% of skills were rated as Poor. Manual inspection of the per-metric \textit{reason} fields indicates that poor executability mainly stems from broken execution paths (\textit{e.g.}, missing required scripts or workflow files, dependencies on undefined external tools or skills) or incomplete specifications (\textit{e.g.}, missing invocation details, incomplete or erroneous code snippets), which prevent the skills from being run as written.
Completeness is the relatively weakest dimension. 55.7\% of skills received a Good rating, while 44.3\% received an Average rating. Further analysis of the \textit{reason} fields shows that Average ratings arise when skills describe the intended workflow but leave reuse-critical details underspecified, including prerequisites and environment setup, concrete commands or scripts, input / output formats, failure handling, and edge case guidance. Overall, SkillNet mainly evaluates \textit{whether a skill is sufficiently specified and executable in its native environment, but cannot validate the generality and transferability} across different SE contexts.

\begin{mdframed}[style=MyFrame]
    {\fontsize{10pt}{0}\selectfont\faStar} \textbf{Finding:} Current skill marketplaces evaluate SE-related skills primarily through marketplace-native metrics that assess \textbf{safety, quality, and readiness}.\\
{\fontsize{10pt}{0}\selectfont\faLightbulbO}
\textbf{Opportunity:} A unified evaluation framework is needed to assess reusable SE activities based on the following properties: \textbf{triggerability} (can the skill be reliably invoked), \textbf{attributability} (can outcomes be attributed to the skill), \textbf{localizability} (can failures be localized), and \textbf{traceability} (can execution be traced), together with their generalizability across SE contexts.
\end{mdframed}

\section{Threats to validity}

We identify several threats to validity in our study and describe the mitigation strategies adopted to reduce their impact. First, \textbf{Internal validity}. A primary threat arises from the use of LLMs to support the annotation of SE-related skills, including identifying the SE activities involved and assigning lifecycle phases. Such annotations may introduce subjectivity or inconsistencies due to model reasoning biases. To mitigate this issue, we employ an advanced model (\textit{i.e.}, Qwen3.6-35B-A3B) to improve annotation reliability and consistency. In addition, we manually validate a representative sample of the annotated data to assess correctness and calibrate for potential annotation errors.
Second, \textbf{External validity}. Our analysis is based on skills collected from public skill marketplaces, which may not fully represent the broader population of agentic skills in the wild. This raises concerns about the generalizability of our findings beyond the selected sources. To mitigate this threat, we follow established practices in prior studies and collect skills from four widely used and diverse skill marketplaces, aiming to improve coverage and representativeness of the dataset. While this does not guarantee full coverage of all existing skills, it provides a broad and practical approximation of the current ecosystem.
Third, \textbf{Construct validity}. Another threat relates to the accuracy of evaluation-related information extracted from skill marketplaces. Specifically, the reported evaluation results and metrics associated with skills may be incomplete, inconsistent, or imprecise, potentially affecting the validity of our analysis of reuse evaluation practices. To address this issue, we retain and provide direct references (\textit{i.e.}, source links) for each skill in our dataset, enabling traceability and verification of all extracted evaluation information against the original marketplace records.

\section{Conclusion}

% In this work, we conducted a large-scale empirical study of SE-related skills in emerging agent marketplaces to understand how SE activities are encapsulated into reusable artifacts. Our analysis shows that these skills provide a new form of reuse for SE knowledge, but their coverage across the software lifecycle is uneven, and their evaluation often does not fully reflect reuse capability. These findings offer an initial activity-centric understanding of SE skills and shed light on how SE reuse is evolving in agent-driven ecosystems.

We conducted a large-scale empirical study of SE-related skills across emerging agent marketplaces to understand how software engineering activities are encapsulated into reusable artifacts. Our findings provide an initial activity-centric view of reusable SE skills, revealing uneven lifecycle coverage and limited reuse-oriented evaluation, and offering insights into the evolution of software engineering reuse in the agent era.

% we presented the first large-scale empirical study of SE-related skills in emerging agentic skill marketplaces. By systematically analyzing a broad corpus of skills, we characterized how SE activities are encapsulated into reusable skill artifacts, and examined their lifecycle coverage, evaluation practices, and composition patterns. Our findings reveal that while skills increasingly serve as a new form of reusable SE knowledge, their coverage across the software lifecycle remains uneven, and current evaluation mechanisms often fail to fully capture their reuse capability. Overall, this study provides an activity-centric understanding of SE skills and offers foundational insights into how software engineering reuse is being reshaped in the era of AI agents.

% \section*{Data Availability}
% The collected skills and code are available at \url{}.

\bibliographystyle{IEEEtran}
\bibliography{refs}

% Generated by IEEEtran.bst, version: 1.14 (2015/08/26)
\begin{thebibliography}{10}
\providecommand{\url}[1]{#1}
\csname url@samestyle\endcsname
\providecommand{\newblock}{\relax}
\providecommand{\bibinfo}[2]{#2}
\providecommand{\BIBentrySTDinterwordspacing}{\spaceskip=0pt\relax}
\providecommand{\BIBentryALTinterwordstretchfactor}{4}
\providecommand{\BIBentryALTinterwordspacing}{\spaceskip=\fontdimen2\font plus
\BIBentryALTinterwordstretchfactor\fontdimen3\font minus \fontdimen4\font\relax}
\providecommand{\BIBforeignlanguage}[2]{{%
\expandafter\ifx\csname l@#1\endcsname\relax
\typeout{** WARNING: IEEEtran.bst: No hyphenation pattern has been}%
\typeout{** loaded for the language `#1'. Using the pattern for}%
\typeout{** the default language instead.}%
\else
\language=\csname l@#1\endcsname
\fi
#2}}
\providecommand{\BIBdecl}{\relax}
\BIBdecl

\bibitem{krueger1992software}
C.~W. Krueger, ``Software reuse,'' \emph{ACM Computing Surveys (CSUR)}, vol.~24, no.~2, pp. 131--183, 1992.

\bibitem{frakes2005software}
W.~B. Frakes and K.~Kang, ``Software reuse research: Status and future,'' \emph{IEEE transactions on Software Engineering}, vol.~31, no.~7, pp. 529--536, 2005.

\bibitem{papazoglou2003service}
M.~P. Papazoglou, ``Service-oriented computing: Concepts, characteristics and directions,'' in \emph{Proceedings of the Fourth International Conference on Web Information Systems Engineering, 2003. WISE 2003.}\hskip 1em plus 0.5em minus 0.4em\relax IEEE, 2003, pp. 3--12.

\bibitem{wang2005component}
A.~J.~A. Wang and K.~Qian, \emph{Component-oriented programming}.\hskip 1em plus 0.5em minus 0.4em\relax John Wiley \& Sons, 2005.

\bibitem{tang2025empowering}
Y.~Tang, K.~Chen, L.~Yue, J.~Fan, C.~Zhou, X.~Li, Y.~Zhang, M.~Zhao, S.~Kai, K.~Guo \emph{et~al.}, ``Empowering real-world: A survey on the technology, practice, and evaluation of llm-driven industry agents,'' \emph{arXiv preprint arXiv:2510.17491}, 2025.

\bibitem{wang2023voyager}
G.~Wang, Y.~Xie, Y.~Jiang, A.~Mandlekar, C.~Xiao, Y.~Zhu, L.~Fan, and A.~Anandkumar, ``Voyager: An open-ended embodied agent with large language models,'' \emph{arXiv preprint arXiv:2305.16291}, 2023.

\bibitem{jiang2026sok}
Y.~Jiang, D.~Li, H.~Deng, B.~Ma, X.~Wang, Q.~Wang, and G.~Yu, ``Sok: Agentic skills--beyond tool use in llm agents,'' \emph{arXiv preprint arXiv:2602.20867}, 2026.

\bibitem{hong2024metagpt}
S.~Hong, M.~Zhuge, J.~Chen, X.~Zheng, Y.~Cheng, J.~Wang, C.~Zhang, S.~Yau, Z.~Lin, L.~Zhou \emph{et~al.}, ``Metagpt: Meta programming for a multi-agent collaborative framework,'' in \emph{International Conference on Learning Representations}, vol. 2024, 2024, pp. 23\,247--23\,275.

\bibitem{qian2024chatdev}
C.~Qian, W.~Liu, H.~Liu, N.~Chen, Y.~Dang, J.~Li, C.~Yang, W.~Chen, Y.~Su, X.~Cong \emph{et~al.}, ``Chatdev: Communicative agents for software development,'' in \emph{Proceedings of the 62nd annual meeting of the association for computational linguistics (volume 1: Long papers)}, 2024, pp. 15\,174--15\,186.

\bibitem{awesome-skills}
\BIBentryALTinterwordspacing
VoltAgent, ``Awesome-agent-skills,'' 2025. [Online]. Available: \url{https://github.com/VoltAgent/awesome-agent-skills}
\BIBentrySTDinterwordspacing

\bibitem{anthropic-skills}
\BIBentryALTinterwordspacing
Anthropics, ``Skills,'' 2025. [Online]. Available: \url{https://github.com/anthropics/skills}
\BIBentrySTDinterwordspacing

\bibitem{liang2026skillnet}
Y.~Liang, R.~Zhong, H.~Xu, C.~Jiang, Y.~Zhong, R.~Fang, J.-C. Gu, S.~Deng, Y.~Yao, M.~Wang \emph{et~al.}, ``Skillnet: Create, evaluate, and connect ai skills,'' \emph{arXiv preprint arXiv:2603.04448}, 2026.

\bibitem{clawhub2026}
\BIBentryALTinterwordspacing
ClawHub, ``Clawhub,'' 2026. [Online]. Available: \url{https://clawhub.ai}
\BIBentrySTDinterwordspacing

\bibitem{skillhub2025}
\BIBentryALTinterwordspacing
SkillHub, ``Skillhub,'' 2025. [Online]. Available: \url{https://www.skillhub.club/}
\BIBentrySTDinterwordspacing

\bibitem{skillnet2026}
\BIBentryALTinterwordspacing
ZJUNLP, ``Skillnet,'' 2026. [Online]. Available: \url{http://skillnet.openkg.cn/}
\BIBentrySTDinterwordspacing

\bibitem{skillsmp2025}
\BIBentryALTinterwordspacing
SkillsMP, ``Skillsmp,'' 2025. [Online]. Available: \url{https://skillsmp.com/}
\BIBentrySTDinterwordspacing

\bibitem{huang2023agentcoder}
D.~Huang, J.~M. Zhang, M.~Luck, Q.~Bu, Y.~Qing, and H.~Cui, ``Agentcoder: Multi-agent-based code generation with iterative testing and optimisation,'' \emph{arXiv preprint arXiv:2312.13010}, 2023.

\bibitem{harman2025mutation}
M.~Harman, J.~Ritchey, I.~Harper, S.~Sengupta, K.~Mao, A.~Gulati, C.~Foster, and H.~Robert, ``Mutation-guided llm-based test generation at meta,'' in \emph{Proceedings of the 33rd ACM International Conference on the Foundations of Software Engineering}, 2025, pp. 180--191.

\bibitem{yang2024swe}
J.~Yang, C.~E. Jimenez, A.~Wettig, K.~Lieret, S.~Yao, K.~Narasimhan, and O.~Press, ``Swe-agent: Agent-computer interfaces enable automated software engineering,'' \emph{Advances in Neural Information Processing Systems}, vol.~37, pp. 50\,528--50\,652, 2024.

\bibitem{yao2022react}
S.~Yao, J.~Zhao, D.~Yu, N.~Du, I.~Shafran, K.~Narasimhan, and Y.~Cao, ``React: Synergizing reasoning and acting in language models,'' \emph{arXiv preprint arXiv:2210.03629}, 2022.

\bibitem{schick2023toolformer}
T.~Schick, J.~Dwivedi-Yu, R.~Dess{\`\i}, R.~Raileanu, M.~Lomeli, E.~Hambro, L.~Zettlemoyer, N.~Cancedda, and T.~Scialom, ``Toolformer: Language models can teach themselves to use tools,'' \emph{Advances in neural information processing systems}, vol.~36, pp. 68\,539--68\,551, 2023.

\bibitem{ni2026trace2skill}
J.~Ni, Y.~Liu, X.~Liu, Y.~Sun, M.~Zhou, P.~Cheng, D.~Wang, E.~Zhao, X.~Jiang, and G.~Jiang, ``Trace2skill: Distill trajectory-local lessons into transferable agent skills,'' \emph{arXiv preprint arXiv:2603.25158}, 2026.

\bibitem{wang2023describe}
Z.~Wang, S.~Cai, G.~Chen, A.~Liu, X.~S. Ma, and Y.~Liang, ``Describe, explain, plan and select: interactive planning with llms enables open-world multi-task agents,'' \emph{Advances in Neural Information Processing Systems}, vol.~36, pp. 34\,153--34\,189, 2023.

\bibitem{li2026skill}
Z.~Li, D.~Guodong, Z.~Shi, W.~Guo, W.~Yao, Y.~Zhou, J.~Zhang, and J.~Li, ``Skill weaving: Efficient llm improvement via modular skillpacks,'' in \emph{Findings of the Association for Computational Linguistics: ACL 2026}, 2026, pp. 40\,000--40\,023.

\bibitem{wang2026skillx}
C.~Wang, Z.~Yu, X.~Xie, W.~Yao, R.~Fang, S.~Qiao, K.~Cao, G.~Zheng, X.~Qi, P.~Zhang \emph{et~al.}, ``Skillx: Automatically constructing skill knowledge bases for agents,'' \emph{arXiv preprint arXiv:2604.04804}, 2026.

\bibitem{bouzenia2025understanding}
I.~Bouzenia and M.~Pradel, ``Understanding software engineering agents: A study of thought-action-result trajectories,'' in \emph{2025 40th IEEE/ACM International Conference on Automated Software Engineering (ASE)}.\hskip 1em plus 0.5em minus 0.4em\relax IEEE, 2025, pp. 2846--2857.

\bibitem{zhou2026comprehensive}
Y.~Zhou, W.~Shu, Y.~Su, W.~Du, Y.~Fang, and X.~Lin, ``A comprehensive survey on agent skills: Taxonomy, techniques, and applications,'' \emph{arXiv preprint arXiv:2605.07358}, 2026.

\bibitem{li2026can}
D.~Li, Y.~He, Y.~Hu, Y.~Tian, and J.~Li, ``Can llm agents generate real-world evidence? evaluating observational studies in medical databases,'' \emph{arXiv preprint arXiv:2603.22767}, 2026.

\bibitem{yu2024fincon}
Y.~Yu, Z.~Yao, H.~Li, Z.~Deng, Y.~Jiang, Y.~Cao, Z.~Chen, J.~W. Suchow, Z.~Cui, R.~Liu \emph{et~al.}, ``Fincon: A synthesized llm multi-agent system with conceptual verbal reinforcement for enhanced financial decision making,'' \emph{Advances in Neural Information Processing Systems}, vol.~37, pp. 137\,010--137\,045, 2024.

\bibitem{ling2026agent}
G.~Ling, S.~Zhong, and R.~Huang, ``Agent skills: A data-driven analysis of claude skills for extending large language model functionality,'' \emph{arXiv preprint arXiv:2602.08004}, 2026.

\bibitem{xu2026agent}
R.~Xu and Y.~Yan, ``Agent skills for large language models: Architecture, acquisition, security, and the path forward,'' \emph{arXiv preprint arXiv:2602.12430}, 2026.

\bibitem{liu2026well}
Y.~Liu, J.~Ji, L.~An, T.~Jaakkola, Y.~Zhang, and S.~Chang, ``How well do agentic skills work in the wild: Benchmarking llm skill usage in realistic settings,'' \emph{arXiv preprint arXiv:2604.04323}, 2026.

\bibitem{ding2026agent}
K.~Ding, Y.~Zhou, C.~Jin, F.~Tong, M.~Zhou, and D.~N. Metaxas, ``Agent skill evaluation and evolution: Frameworks and benchmarks,'' \emph{arXiv preprint arXiv:2606.11435}, 2026.

\bibitem{zeng2025benchmarking}
Z.~Zeng, Y.~Li, R.~Xie, W.~Ye, and S.~Zhang, ``Benchmarking and studying the llm-based agent system in end-to-end software development,'' \emph{arXiv preprint arXiv:2511.04064}, 2025.

\bibitem{zhang2026coevoskills}
H.~Zhang, S.~Fan, H.~P. Zou, Y.~Chen, Z.~Wang, J.~Zhou, C.~Li, W.-C. Huang, Y.~Yao, K.~Zheng \emph{et~al.}, ``Coevoskills: Self-evolving agent skills via co-evolutionary verification,'' \emph{arXiv preprint arXiv:2604.01687}, 2026.

\bibitem{makitalo2020opportunistic}
N.~M{\"a}kitalo, A.~Taivalsaari, A.~Kiviluoto, T.~Mikkonen, and R.~Capilla, ``On opportunistic software reuse: N. m{\"a}kitalo et al.'' \emph{Computing}, vol. 102, no.~11, pp. 2385--2408, 2020.

\bibitem{capilla2019opportunities}
R.~Capilla, B.~Gallina, C.~Cetina, and J.~Favaro, ``Opportunities for software reuse in an uncertain world: From past to emerging trends,'' \emph{Journal of software: Evolution and process}, vol.~31, no.~8, p. e2217, 2019.

\bibitem{cwl2021}
\BIBentryALTinterwordspacing
M.~R. Crusoe, S.~Abeln, A.~Iosup, P.~Amstutz, J.~Chilton, N.~Tijanic, H.~M{\'{e}}nager, S.~Soiland{-}Reyes, and C.~A. Goble, ``Methods included: Standardizing computational reuse and portability with the common workflow language,'' \emph{CoRR}, vol. abs/2105.07028, 2021. [Online]. Available: \url{https://arxiv.org/abs/2105.07028}
\BIBentrySTDinterwordspacing

\bibitem{workflow2025}
\BIBentryALTinterwordspacing
M.~S. Nikoo, S.~Kochanthara, {\"{O}}.~Babur, and M.~van~den Brand, ``An empirical study of business process models and model clones on github,'' \emph{Empir. Softw. Eng.}, vol.~30, no.~2, p.~48, 2025. [Online]. Available: \url{https://doi.org/10.1007/s10664-024-10584-z}
\BIBentrySTDinterwordspacing

\bibitem{moreira2022open}
R.~A.~F. Moreira, W.~K. Assun{\c{c}}{\~a}o, J.~Martinez, and E.~Figueiredo, ``Open-source software product line extraction processes: the argouml-spl and phaser cases,'' \emph{Empirical Software Engineering}, vol.~27, no.~4, p.~85, 2022.

\bibitem{jung2024automated}
P.~Jung, S.~Lee, and U.~Lee, ``Automated code-based test case reuse for software product line testing,'' \emph{Information and Software Technology}, vol. 166, p. 107372, 2024.

\bibitem{gulwani2017programsynthesis}
\BIBentryALTinterwordspacing
S.~Gulwani, O.~Polozov, and R.~Singh, ``Program synthesis,'' \emph{Found. Trends Program. Lang.}, vol.~4, no. 1-2, pp. 1--119, 2017. [Online]. Available: \url{https://doi.org/10.1561/2500000010}
\BIBentrySTDinterwordspacing

\bibitem{liu2019accelerating}
B.~Liu, W.~Dong, and Y.~Zhang, ``Accelerating api-based program synthesis via api usage pattern mining,'' \emph{IEEE Access}, vol.~7, pp. 159\,162--159\,176, 2019.

\bibitem{reuse-in-ai-era2025}
\BIBentryALTinterwordspacing
T.~Mikkonen and A.~Taivalsaari, ``Software reuse in the generative {AI} era: From cargo cult towards {AI} native software engineering,'' \emph{CoRR}, vol. abs/2506.17937, 2025. [Online]. Available: \url{https://doi.org/10.48550/arXiv.2506.17937}
\BIBentrySTDinterwordspacing

\bibitem{liu2026agent}
Y.~Liu, W.~Wang, R.~Feng, Y.~Zhang, G.~Xu, G.~Deng, Y.~Li, and L.~Zhang, ``Agent skills in the wild: An empirical study of security vulnerabilities at scale,'' \emph{arXiv preprint arXiv:2601.10338}, 2026.

\bibitem{agenticOpportunity25}
T.~Raheem and G.~Hossain, ``Agentic ai systems: Opportunities, challenges, and trustworthiness,'' in \emph{2025 IEEE International Conference on Electro Information Technology (eIT)}, 2025, pp. 618--624.

\bibitem{li2026organizing}
H.~Li, C.~Mu, J.~Chen, S.~Ren, Z.~Cui, Y.~Zhang, L.~Bai, and S.~Hu, ``Organizing, orchestrating, and benchmarking agent skills at ecosystem scale,'' \emph{arXiv preprint arXiv:2603.02176}, 2026.

\bibitem{openclaw2026}
\BIBentryALTinterwordspacing
{OpenClaw}, ``Openclaw: The personal ai assistant framework,'' 2026. [Online]. Available: \url{https://openclaw.ai/}
\BIBentrySTDinterwordspacing

\bibitem{openai2026gpt55}
\BIBentryALTinterwordspacing
OpenAI, ``Gpt-5.5 system card,'' 2026. [Online]. Available: \url{https://openai.com/index/gpt-5-5-system-card/}
\BIBentrySTDinterwordspacing

\bibitem{openai2026tiktoken}
\BIBentryALTinterwordspacing
{OpenAI}, ``Tiktoken: A fast bpe tokeniser for use with openai's models,'' 2026. [Online]. Available: \url{https://github.com/openai/tiktoken}
\BIBentrySTDinterwordspacing

\bibitem{qwen36_35b_a3b}
\BIBentryALTinterwordspacing
{Qwen Team}, ``{Qwen3.6-35B-A3B}: Agentic coding power, now open to all,'' April 2026. [Online]. Available: \url{https://qwen.ai/blog?id=qwen3.6-35b-a3b}
\BIBentrySTDinterwordspacing

\bibitem{openclaw_security_audits}
\BIBentryALTinterwordspacing
OpenClaw, ``Security audits,'' 2026. [Online]. Available: \url{https://docs.openclaw.ai/clawhub/security-audits}
\BIBentrySTDinterwordspacing

\bibitem{nvidia_skillspector}
\BIBentryALTinterwordspacing
NVIDIA, ``Skillspector: Security scanner for ai agent skills,'' 2026. [Online]. Available: \url{https://github.com/NVIDIA/SkillSpector}
\BIBentrySTDinterwordspacing

\bibitem{openclaw_security_prompt}
\BIBentryALTinterwordspacing
OpenClaw, ``securityprompt.ts,'' 2026. [Online]. Available: \url{https://github.com/openclaw/clawhub/blob/main/convex/lib/securityPrompt.ts}
\BIBentrySTDinterwordspacing

\bibitem{virustotal}
\BIBentryALTinterwordspacing
VirusTotal, ``Virustotal,'' 2026. [Online]. Available: \url{https://www.virustotal.com/}
\BIBentrySTDinterwordspacing

\end{thebibliography}

\end{document}